\documentclass[prb, aps, reprint, superscriptaddress, floatfix, showpacs, showkeys]{revtex4-2}
\usepackage{amsmath}
\usepackage{amssymb}
\usepackage{bm}
\usepackage{graphicx}
\graphicspath{{./figures/}}

\usepackage[svgnames]{xcolor}

\newcommand{\mvec}[1]{\ensuremath{\mathbf{#1}}} 
 
\newcommand{\Heff}{\ensuremath{\mathbf{H}_\mathrm{eff}}}

\renewcommand{\thetable}{\arabic{table}}

\makeatletter
\renewcommand*{\fnum@figure}{{\normalfont\bfseries Figure~\thefigure}}
\renewcommand*{\fnum@table}{{\normalfont\bfseries Listing~\thetable}}
\makeatother

\def\mean#1{\left< #1 \right>}

\begin{document}
\title{MicroMagnetic.jl: A Julia package for micromagnetic and atomistic simulations with GPU support}

\date{\today}

\author{Weiwei Wang}
\email{wangweiwei@ahu.edu.cn}
\affiliation{Institutes of Physical Science and Information Technology, Anhui University, Hefei 230601, China}

\author{Boyao Lyu}
\affiliation{Anhui Province Key Laboratory of Low-Energy Quantum Materials and Devices, High Magnetic Field Laboratory, HFIPS, Chinese Academy of Sciences, Hefei, Anhui 230031, China}       
\affiliation{University of Science and Technology of China, Hefei 230031, China}

\author{Lingyao Kong}
\affiliation{School of Physics and Optoelectronic Engineering, Anhui University, Hefei, 230601, China}

\author{Hans Fangohr}
\affiliation{University of Southampton, Southampton SO17 1BJ, United Kingdom}

\author{Haifeng Du}
\email{duhf@hmfl.ac.cn}
\affiliation{Anhui Province Key Laboratory of Low-Energy Quantum Materials and Devices, High Magnetic Field Laboratory, HFIPS, Chinese Academy of Sciences, Hefei, Anhui 230031, China}

\begin{abstract}
MicroMagnetic.jl is an open-source Julia package for micromagnetic and atomistic simulations. Using the features of the Julia programming language, 
MicroMagnetic.jl supports CPU and various GPU platforms, including NVIDIA, AMD, Intel, and Apple GPUs. Moreover, MicroMagnetic.jl supports Monte Carlo simulations 
for atomistic models and implements the Nudged-Elastic-Band method for energy barrier computations. With built-in support for double and 
single precision modes and a design allowing easy extensibility to add new features, MicroMagnetic.jl provides a versatile toolset for researchers 
in micromagnetics and atomistic simulations.
\end{abstract}


\maketitle

\section{Introduction}

Micromagnetics~\cite{Kronmuller2007, Leliaert2019} is an effective tool for studying magnetic structures 
and phenomena at the micrometer scale. With the fast development of high-performance computation,  micromagnetic simulations 
have been applied in various technological applications, such as magnetic recording media, magnonics~\cite{Rezende2020, Flebus2024}, magnetic sensors, 
and spintronic devices~\cite{Grollier2020, Abert2019, Barla2021}. 

Micromagnetic simulation packages typically use two primary numerical methods: finite difference methods (FDM)~\cite{Miltat2007} 
and finite element methods (FEM)~\cite{Schrefl2007}. 
The FDM approach, as implemented in packages such as OOMMF~\cite{Porter1999}, MuMax3~\cite{Vansteenkiste2014}, and fidimag~\cite{Bisotti2018},  
offers simplicity in implementation and computational efficiency. The FEM, as employed in Magpar~\cite{Scholz2003}, Nmag~\cite{Fischbacher2007} and Commics~\cite{Pfeiler2020}, 
provides more flexibility in handling complex geometries at the cost of increased computational complexity and setup efforts.

The rapid development of graphics processing units (GPUs), especially Nvidia's GPUs, has extensively promoted the development of high-performance computing. 
GPUs' parallelization capabilities have accelerated micromagnetic simulations. For example, MuMax3~\cite{Vansteenkiste2014} uses Nvidia GPUs to speed up 
the micromagnetic calculations.

The micromagnetic packages fidimag~\cite{Bisotti2018} and magnum.np~\cite{Bruckner2023} are developed using the high-level language Python, 
which provides a uniform post-processing toolchain. 
Similarly, Julia's design prioritizes high performance and ease of use, making it well-suited for scientific tasks. 
Moreover, JuliaGPU~\cite{Besard2019} reuses Julia's existing CPU-oriented compiler to generate code for the GPU, allowing a subset of the language to be directly executable on the GPU.
JuliaGPU initially supported only NVIDIA hardware accelerators via CUDA. However, with GPUCompiler.jl and LLVM, 
Julia's capabilities extend to various hardware platforms, including AMD, Intel, and Apple M-series GPUs.

MicroMagnetic.jl is an open-source Julia package developed under the MIT license and designed for micromagnetic and atomistic simulations. 
MicroMagnetic.jl can be installed directly using Julia's package manager \texttt{Pkg}.  
Using the KernelAbstractions.jl~\cite{ValentinChuravy2024} library, MicroMagnetic.jl facilitates the creation of heterogeneous 
kernels for various backends within a unified framework.

  MicroMagnetic.jl is built on a modular framework, allowing others in the micromagnetic community to easily integrate new physics modules. 
  Currently, the micromagnetic module in MicroMagnetic.jl supports only the finite difference method. In the future, we plan to incorporate a 
  finite element implementation. Additionally, we will expand the atomistic module to include more lattice structures and enable direct parameter 
  input from first-principles calculations. Finally, we are preparing to introduce a visualization interface to allow users to view simulation results in real-time.

\section{Design}

\begin{figure}[htbp]
  \begin{center}
  \includegraphics[width=0.4\textwidth]{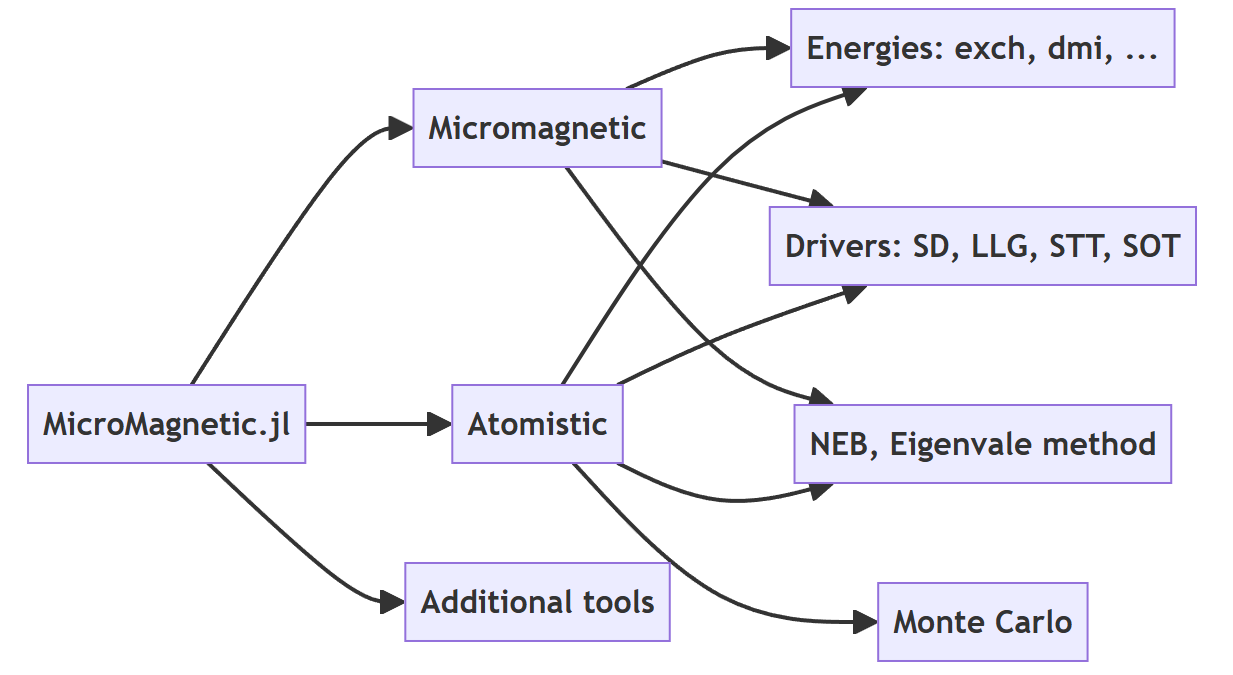} 
  \caption{The basic structure of MicroMagnetic.jl.}
  \label{fig1}
  \end{center}
\end{figure}

Multiple dispatch in Julia enables dynamic method selection based on argument types, enhancing code flexibility and extensibility. 
In MicroMagnetic.jl, we define various types to make use of multiple dispatch. For example, the \texttt{FDMesh} stores the discretized grid information 
since we have used FDM to discretize the micromagnetic energies. Fig.~\ref{fig1} shows the basic structure of MicroMagnetic.jl, which mainly contains three 
parts: standard micromagnetics, atomistic simulations, and additional tools like LETM, MFM and post-processing tools.

\begin{table}\label{tab1}
  \begin{verbatim}
  function circular_Ms(x, y, z)
    if x^2 + y^2 <= (50nm)^2
        return 8.0e5
    end
    return 0.0
  end
  
  set_Ms(sim, circular_Ms)
\end{verbatim}
\caption{An example of the function-based interface: users can pass their own defined functions to MicroMagnetic.jl.}
\end{table}

The micromagnetic parameters provided by the user include material properties, discretization details, and sample shapes. 
To provide users with maximum flexibility, almost all set functions in MicroMagnetic.jl can accept functions as parameters, as designed in Nmag~\cite{Fischbacher2007}, 
FinMag~\cite{Marc-AntonioBisotti2018} and fidimag~\cite{Bisotti2018}. 
For instance, the \texttt{set\_Ms} function can be used to set the system's saturation magnetization. Moreover, it also can be used to define shapes.
In Listing~1, we have defined a round disk, beyond which its saturation magnetization is 0. 
This cell-based approach enables users to define spatial parameters, offering maximum flexibility.

\begin{figure}[htbp]
  \begin{center}
  \includegraphics[width=0.45\textwidth]{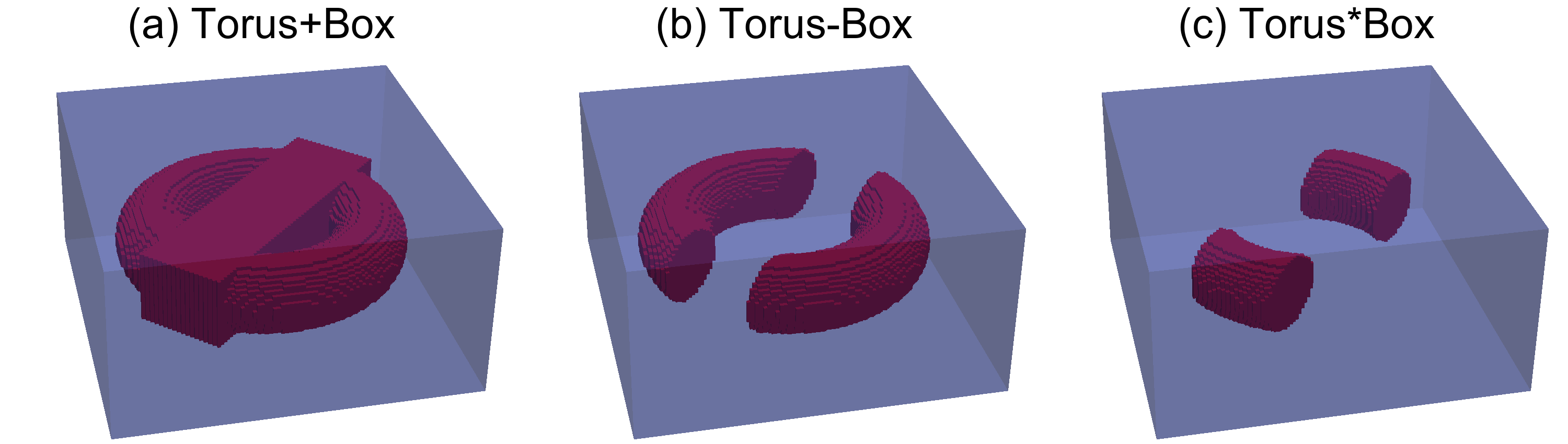} 
  \caption{The boolean operations of a torus and a rotated box: (a) The union '$+$', (b) The difference '$-$', and (c) The intersection '$*$'.}
  \label{fig2}
  \end{center}
\end{figure}

In addition to defining shapes using \texttt{set\_Ms} function, MicroMagnetic.jl provides basic shapes and boolean operations to create regular shapes and their combinations,
similar to MuMax3. 
Supported basic shapes in MicroMagnetic.jl include planes, cylinders, spheres, boxes, and torus. Fig.\ref{fig2} shows the boolean operations, including the 
union '$+$', the difference '$-$' and the intersection '$*$' between a torus and a rotated box.

\begin{table}\label{tab2}
\begin{verbatim}
using MicroMagnetic

args = (
    task="relax", 
    mesh=FDMesh(nx=4, ny=4, nz=4), 
    Ms=1e6,
    A=1e-11,
    Ku=5e4,
    axis=(1, 1, 0),
    stopping_dmdt=0.05,
    H_s=[(i * 5mT, 0, 0) for i in -20:20]
)
  
sim_with(args)
\end{verbatim}
\caption{The Stoner-Wohlfarth model using the high-level interface sim\_with.}
\end{table}

  Beyond the low-level API, which includes functions like \texttt{set\_Ms}, we provide a high-level interface called \texttt{sim\_with} 
  to simplify the setup and execution of micromagnetic simulations. This interface allows users to encapsulate all relevant micromagnetic 
  parameters into a \texttt{NamedTuple} or \texttt{Dict}, which can then be passed directly to \texttt{sim\_with}.  Listing~2 illustrates 
  the use of \texttt{sim\_with} for calculating the Stoner-Wohlfarth model. In this example, the external field $H$
  is varied using the \texttt{\_sweep} suffix (or \texttt{\_s} for short). This \texttt{\_s} syntax can similarly applied to other parameters, 
  such as \texttt{Ms}, \texttt{Ku}, \texttt{A}, \texttt{D}, \texttt{task}, and \texttt{driver}, enabling the simulation to iterate over different 
  values for these parameters. This approach provides the flexibility needed to explore a wide range of micromagnetic scenarios, 
  including the computation of hysteresis loops and the investigation of parameter-dependent phenomena.

\section{Micromagnetics} 
In the continuum theory, the magnetization $\mathbf{M}(\mathbf{r})$ is used to describe the average moment density over a local volume, 
and the total micromagnetic energy $E$ is a function of the magnetization  $\mathbf{M}$.
The effective field $\Heff$ is defined as the functional derivative~\cite{Kronmuller2007}
\begin{equation}
\Heff=-\frac{1}{\mu_0 M_s} \frac{\delta E}{\delta \mathbf{m}},
\end{equation}
where  $\mathbf{m}=\mathbf{M}(\mathbf{r})/M_s$ is the unit vector of the magnetization with $M_s$ the saturation magnetization. 

\subsection{Exchange energy}
In the continuum limit, the isotropic exchange energy can be written as~\cite{Miltat2007}
\begin{gather}
\begin{split}E_\mathrm{ex} = \int_\Omega A (\nabla \mathbf{m})^2 dx \end{split}
\end{gather}
where $(\nabla \mathbf{m})^2 = (\nabla m_x)^2 + (\nabla m_y)^2 + (\nabla m_z)^2$. 
Therefore, the effective field for exchange interaction is 
\begin{equation}
  \mathbf{H}_\mathrm{ex}=\frac{2 A}{\mu_0 M_s} \nabla^2 \mathbf{m}.
\end{equation}
Taking into account the standard 6-neighbors, the exchange energy density of cell $i$ reads~\cite{Porter1999}
\begin{equation}
  w_i =\sum_{j \in N_i} A_{i j} \frac{\mathbf{m}_i \cdot\left(\mathbf{m}_i-\mathbf{m}_j\right)}{\Delta_{i j}^2}
\end{equation}
where $N_i$ denotes the set of 6 neighbors, $A_{ij}$ represents the exchange coefficient between cells $i$ and $j$, 
and $\Delta_{i j}$ is discretization step size. $A_{i j}=A$ for the uniform exchange constant. MicroMagnetic.jl also supports the spatial exchange constant, 
i.e., cell-based exchange constant $A_i$ for cell $i$. Under this scenario $A_{i j}=2 A_i A_j/(A_i+A_j)$~\cite{Porter1999}.
Consequently, the corresponding effective field can be computed as $\mathbf{H}_\mathrm{ex, i} = -1/(\mu_0 M_s) (\partial w_i / \partial \mathbf{m}_i)$, yielding
\begin{equation}
  \mathbf{H}_\mathrm{ex, i} = \frac{2 }{\mu_0 M_s} \sum_{j \in N_i} A_{i j} \frac{\mathbf{m}_j-\mathbf{m}_i}{\Delta_{i j}^2}.
\end{equation}

\subsection{Magnetostatic energy}
In micromagnetics, the magnetostatic energy also referred to as demagnetization energy, is defined as
\begin{equation}\label{eq_demag_energy}
E_{\mathrm{d}}=
-\frac{\mu_0}{2} \int_V \mathbf{H}_{d}(\mathbf{r}) \cdot \mathbf{M}(\mathbf{r}) dV.
\end{equation}
where $\mathbf{H}_{d}$ denotes the demagnetizing field. Employing FDM, the demagnetization energy can be approximated as~\cite{Miltat2007}
\begin{equation}\label{eq_demag_energy2}
  E_{\mathrm{d}} \approx \frac{\mu_0}{2} \sum_{i,j} |V_i| \mathbf{M}(\mathbf{r}_i)\cdot \mathcal{N}(\mathbf{r}_i,\mathbf{r}_j) \cdot \mathbf{M}(\mathbf{r}_j),
\end{equation}
where $|V_i|= \Delta_x \Delta_y \Delta_z$ represents the cell size, and the demagnetization tensor $\mathcal{N}(\mvec{r}_i,\mvec{r}_j)$
is a $3\times 3$ matrix~\cite{Newell1993}. This tensor exhibits symmetry properties and is solely dependent on the distance between $\mvec{r}_i$ and $\mvec{r}_j$, 
denoted as $\mathcal{N}_{i-j}=\mathcal{N}(\mvec{r}_i-\mvec{r}_j)=\mathcal{N}(\mvec{r}_i,\mvec{r}_j)$.
Consequently, one arrives at a discrete convolution for the demagnetizing field
\begin{equation}\label{eq_hd1}
\mathbf{H}_{\mathrm{d},i} =-\sum_j \mathcal{N}_{i-j}\mathbf{M}_j,
\end{equation}
which represents a cell-average field. 
The computation of the demagnetizing field can be sped up using FFT techniques~\cite{Abert2015}. 
The details of these techniques are clearly illustrated through discrete Fourier transforms~\cite{Wang2015}. 
Each demagnetization field calculation typically requires three forward Fourier transforms and three inverse Fourier transforms. 
However, by using the magnetic scalar potential, the three inverse Fourier transforms can be reduced to one~\cite{Abert2012}.

\subsection{Dzyaloshinskii-Moriya Energy} 
In the continuum limit, the DMI energy density $w_\mathrm{dmi}$ is associated with the so-called \textit{Lifshitz invariants}, which are terms in the form
\begin{equation}
L^{(k)}_{ij} = m_i \frac{\partial m_j}{\partial x_k} - m_j \frac{\partial m_i}{\partial x_k}.
\end{equation}
The form of DMI energy density varies depending on the symmetry class. For bulk DMI found in materials such as MnSi~\cite{Muhlbauer2009} and FeGe~\cite{Huang2012}, 
corresponding to symmetry class $T$ or $O$, the expression is given by~\cite{Bruckner2023, Cortes-Ortuno2018}: 
\begin{equation}
  w_\mathrm{dmi} = D(L^{(z)}_{yx} + L^{(y)}_{xz} + L^{(x)}_{zy}) = D \mathbf{m} \cdot (\nabla \times \mathbf{m}).
\end{equation}
The associated effective field is
\begin{gather}
  \mathbf{H}_\mathrm{dmi}=-\frac{2D}{\mu_0 M_s} (\nabla \times \mathbf{m}).
\end{gather}
For a thin film with interfacial DMI or a crystal with symmetry class $C_{nv}$, the energy density is~\cite{Cortes-Ortuno2018}
\begin{equation}
  w_\mathrm{dmi}=D (L_{x z}^{(x)}+L_{y z}^{(y)} )=D\left(\mathbf{m} \cdot \boldsymbol{\nabla} m_z-m_z \boldsymbol{\nabla} \cdot \mathbf{m}\right),
\end{equation}
and the effective field is 
\begin{equation}
  \mathbf{H}_\mathrm{dmi}=-\frac{2 D}{\mu_0 M_s} (\mathbf{e}_y \times \frac{\partial \mathbf{m}}{\partial x} - \mathbf{e}_x \times \frac{\partial \mathbf{m}}{\partial y}).
\end{equation}
For a crystal with symmetry class $D_{2d}$, the DMI energy density is given by~\cite{Cortes-Ortuno2018} $w_{\mathrm{dmi}}=D (L_{x z}^{(y)}+L_{y z}^{(x)})$, resulting in the effective field
\begin{equation}
  \mathbf{H}_\mathrm{dmi}=-\frac{2 D}{\mu_0 M_s} (\mathbf{e}_y \times \frac{\partial \mathbf{m}}{\partial y} - \mathbf{e}_x \times \frac{\partial \mathbf{m}}{\partial x} ).
\end{equation}
Although the effective fields for different symmetries differ, the numerical implementation can be unified as follows
\begin{equation}
  \mathbf{H}_\mathrm{dmi, i} = -\frac{1}{\mu_0 M_s} \sum_{j \in N_i} D_{ij} \frac{\mathbf{e}_{ij} \times \mathbf{m}_j}{\Delta_{i j}},
\end{equation}
where $D_{ij}$ represents the effective DMI constant and $\mathbf{e}_{ij}$ denotes the DMI vectors.
For bulk DMI, $\mathbf{e}_{ij} = \hat{\mathbf{r}}_{ij}$ where $\hat{\mathbf{r}}_{ij}$ is the unit vector between cell $i$ and cell $j$.
For interfacial DMI,  $\mathbf{e}_{ij} = \mathbf{e}_z \times \mathbf{\hat{r}}_{ij}$, i.e., $\mathbf{e}_{ij}=\{-\mathbf{e}_y, \mathbf{e}_y, \mathbf{e}_x, -\mathbf{e}_x, 0, 0\}$ 
for the 6 neighbors $N_{i}=\{-x,+x,-y,+y,-z, +z\}$. 
For the symmetry class $D_{2d}$ one has $\mathbf{e}_{ij}=\{\mathbf{e}_x, -\mathbf{e}_x, -\mathbf{e}_y, \mathbf{e}_y, 0, 0\}$.
If the cell-based DMI is provided, the effective DMI constant can be computed as $D_{i j}=2 D_i D_j/(D_i+D_j)$~\cite{Bruckner2023}.

\subsection{Interlayer Exchange Interaction}
In multilayer systems, indirect exchange interactions across ferromagnetic layers can occur in addition to the intralayer exchange interactions. 
The interlayer exchange interaction can be either symmetric (RKKY-type) or antisymmetric (DMI-type)~\cite{Vedmedenko2019, Han2019}.
The energy density for the RKKY interaction is given by:
\begin{equation}
w_\mathrm{rkky} = - J_\mathrm{rkky} \mathbf{m}_{i} \cdot \mathbf{m}_{j}
\end{equation}
where $J_\mathrm{rkky}$ is the coupling constant between layer $i$ and layer $j$. The sign of $J_\mathrm{rkky}$ depends on the thickness of the spacer layer.
Similarly, the energy density for the interlayer DMI is:
\begin{equation}
w_\mathrm{dmi-int} = \mathbf{D}_\mathrm{int} \cdot \left(\mathbf{m}_{i} \times \mathbf{m}_{j} \right)
\end{equation}
where $\mathbf{D}_\mathrm{int}$ is the DMI vector. The effective fields for the RKKY and interlayer DMI interactions are:
\begin{equation}
\mathbf{H}_i = \frac{1}{\mu_0 M_s} \frac{J_\mathrm{rkky}}{\Delta_z} \mathbf{m}_{j}, \quad \mathbf{H}_i = \frac{1}{\mu_0 M_s} \frac{\mathbf{D}_\mathrm{int}}{\Delta_z} \times \mathbf{m}_{j}
\end{equation}
where $\Delta_z$ is the thickness of the cell.

\section{Atomistic Spin Model} 
The fundamental assumption of the atomistic spin model is that each lattice site possesses a magnetic moment denoted by $\mu_s$. 
For metallic systems with quenched orbital moments, this magnetic moment primarily arises from its spin angular momentum, expressed as~\cite{Tatara2008}
\begin{equation}
\bm{\mu} = - g \mu_B \mathbf{S} = - \hbar \gamma \mathbf{S}
\end{equation}
where $\mu_B=e \hbar /(2m)$ denotes the Bohr magneton, with $e(>0)$ the electron charge, $\gamma=g\mu_B/\hbar (>0) $ the gyromagnetic ratio, 
and $g=2$ the g-factor. The magnetic moment can freely rotate in three-dimensional space while maintaining a constant magnitude. 
Hence, the atomistic spin model is also referred to as the classical spin model~\cite{Nowak2007}. Various interactions occur between magnetic moments, 
including exchange interaction $\mathcal{H}_\mathrm{ex}$, anisotropy interaction $\mathcal{H}_\mathrm{an}$, dipolar interaction $\mathcal{H}_d$, 
Dzyaloshinskii-Moriya interaction $\mathcal{H}_\mathrm{dmi}$, and Zeeman interaction $\mathcal{H}_z$. The total Hamiltonian is the summation of these interactions~\cite{Skubic2008},
\begin{gather}
\mathcal{H} = \mathcal{H}_\mathrm{ex} + \mathcal{H}_\mathrm{dmi} + \mathcal{H}_d + \mathcal{H}_\mathrm{an} + \mathcal{H}_z,
\end{gather}
The effective field $\Heff$ can be computed from the total Hamiltonian $\mathcal{H}$
\begin{gather}\label{eq_heff}
\Heff = - \frac{1}{\mu_s} \frac{\partial \mathcal{H}}{\partial \mvec{m}},
\end{gather}
where $\mvec{m}=\bm{\mu}/\mu_s$ is the unit vector of magnetic moment $\bm{\mu}$.
In general, parameters in the atomistic spin model can be established through two methods~\cite{Evans2014}: either via \textit{ab initio} 
density functional theory calculations or determined experimentally.

\subsection{Exchange interaction}
The classical Heisenberg Hamiltonian with nearest-neighbor exchange interaction is expressed as 
\begin{gather}\label{eq_Hex} 
   \mathcal{H}_\mathrm{ex} = -  \sum_{\langle i,j\rangle} J_{ij} \mvec{m}_i \cdot \mvec{m}_j, 
\end{gather} 
where $\langle i,j\rangle$ denotes a unique pair of lattice sites $i$ and $j$, and the summation is performed once for each pair. 
The exchange constant $J_{ij}$ characterizes the strength of the exchange interaction. A positive $J_{ij}$ favors the ferromagnetic state 
because the Hamiltonian [Eq.~(\ref{eq_Hex})] is minimized when the magnetic moments $\mvec{m}_i$ and $\mvec{m}_j$ are parallel, 
while a negative $J_{ij}$ leads to the antiferromagnetic state. The effective exchange field at site $i$ can be computed as 
\begin{gather} 
  \mvec{H}_\mathrm{ex,i} = \frac{1}{\mu_s} \sum_{\langle i,j\rangle} J_{ij}  \mvec{m}_j. 
\end{gather}

\subsection{Dzyaloshinskii-Moriya interaction (DMI)}
In a more general form, the Hamiltonian of Heisenberg exchange [Eq.~(\ref{eq_Hex})] can be extended to~\cite{Evans2014}
\begin{equation}\label{eq_Hex2}
\mathcal{H}_\mathrm{ex} = - \sum_{\mean{i,j}}\mvec{m}_i^T \hat{\mathcal{J}}_{ij} \mvec{m}_j,
\end{equation}
where $\hat{\mathcal{J}}_{ij}=\{ J_{ij}^{\mu\nu}\}$ with $\mu, \nu =x,y,z$ is the exchange tensor. 
The diagonal part $\hat{\mathcal{J}}_{ij}^\mathrm{diag}$ results in the isotropic exchange interaction [Eq.~(\ref{eq_Hex})]. 
Therefore, the exchange tensor $\hat{\mathcal{J}}_{ij}$ can be decomposed into three parts
\begin{equation}
  \hat{\mathcal{J}}_{ij} = \hat{\mathcal{J}}_{ij}^\mathrm{diag} +  \hat{\mathcal{J}}_{ij}^s + \hat{\mathcal{J}}_{ij}^a,
\end{equation}
where the traceless symmetric anisotropic exchange tensor $\hat{\mathcal{A}}_{ij}^s$ is defined by
\begin{equation}
  \hat{\mathcal{J}}_{ij}^s  =  \frac{1}{2}(\hat{\mathcal{J}}_{ij}+\hat{\mathcal{J}}_{ij}^T) - \hat{\mathcal{J}}_{ij}^\mathrm{diag},
\end{equation}
and the antisymmetric exchange matrix tensor is given by 
\begin{equation}
  \hat{\mathcal{J}}_{ij}^a  =  \frac{1}{2}(\hat{\mathcal{J}}_{ij}-\hat{\mathcal{J}}_{ij}^T).
\end{equation}
Hence, the contribution to the Hamiltonian [Eq.~(\ref{eq_Hex2})] from $\hat{\mathcal{J}}_{ij}^a$ can be recast into
\begin{equation}\label{eq_Hdmi}
\mathcal{H}_\mathrm{dmi} = \sum_{\mean{i,j}} \mvec{D}_{ij} \cdot (\mvec{m}_i \times \mvec{m}_j),
\end{equation} 
where $D_{ij}^x = J_{ij}^{zy} - J_{ij}^{yz}$, $D_{ij}^y = J_{ij}^{xz} - J_{ij}^{zx}$, and $D_{ij}^z = J_{ij}^{yx} - J_{ij}^{xy}$.
This antisymmetric exchange interaction was first studied by Dzyaloshinskii (1958)~\cite{Dzyaloshinskii1958}
and Moriya (1960)~\cite{Moriya1960}. In general the DMI can arise from the spin-orbit interaction and the vector 
$\mvec{D}$ lies parallel or perpendicular to the line connecting the two spins. The parallel case $\mvec{D}_{ij} = D \hat{\mvec{r}}_{ij}$ corresponds
to the bulk DMI and the perpendicular case $\mvec{D}_{ij} = D \hat{\mvec{r}}_{ij} \times \mvec{e}_z$ corresponds to the interfacial DMI 
if both spins are located in the $xy$-plane~\cite{Rohart2013}. 
The effective field of DMI can be computed as
\begin{equation}
\mvec{H}_\mathrm{dmi,i} = - \frac{1}{\mu_s} \frac{\partial \mathcal{H}_\mathrm{dmi}} {\partial \mvec{m}_i} 
= \frac{1}{\mu_s}  \sum_{\mean{i,j}} \mvec{D}_{ij}\times\mvec{m}_j.
\end{equation}

\subsection{Dipolar interaction}
The dipolar interaction is a long-range interaction.
The Hamiltonian for dipolar interaction between magnetic moments $\bm{\mu}_i$ and $\bm{\mu}_j$ is
\begin{equation}
\mathcal{H}_\mathrm{d} =-\frac{\mu_0 \mu_s^2}{4\pi}\sum_{i<j}\frac{3 (\mvec{m}_i\cdot \hat{\mvec{r}}_{ij})(\mvec{m}_j\cdot \hat{\mvec{r}}_{ij}) 
-\mvec{m}_i \cdot \mvec{m}_j}{r_{ij}^3}, 
\end{equation}
where $r_{ij}$ is the distance between two magnetic moments. Therefore, the corresponding effective field can be computed by
\begin{equation}\label{eq_dipolar}
\mvec{H}_\mathrm{d,i} =\frac{\mu_0 \mu_s }{4\pi}\sum_{i \neq j}\frac{3 \hat{\mvec{r}}_{ij} (\mvec{m}_j\cdot \hat{\mvec{r}}_{ij}) -  \mvec{m}_j}{r_{ij}^3}.
\end{equation}
Similar to the micromagnetic case, the calculation of dipolar interaction can be sped up using FFT. 

\subsection{Anisotropy}
\label{core_eqs:anisotropy}
In the presence of anisotropy, a magnetic moment tends to align along some preferred direction. Many physical effects can lead to an anisotropy,
for instance, magnetocrystalline anisotropy can arise from the interaction between the local crystal environment and atomic electron orbitals~\cite{Evans2014}.
The simplest form of the anisotropy is the so-called uniaxial anisotropy, and its Hamiltonian is given by  
\begin{gather}
\begin{split}\mathcal{H}_\mathrm{an} = - K \sum_i (\mvec{m}_i \cdot \mvec{u})^2, \end{split}\notag
\end{gather}
where the unit vector $\mvec{u}$ is the easy axis and the constant $K$ represents the anisotropy strength.
The corresponding effective field due to the uniaxial anisotropy is
\begin{gather}
\begin{split}\mvec{H}_\mathrm{an,i} = \frac{2 K}{\mu_s} (\mvec{m}_i \cdot \mvec{u}) \mvec{u}.\end{split}
\end{gather}
Some materials, such as Nickel, have a cubic crystal structure, which gives them a different form of anisotropy called cubic anisotropy. 
The Hamiltonian of cubic anisotropy is given by
\begin{equation}\label{eq_cubic_ham}
\mathcal{H}_\mathrm{an}^c = K_c \sum_i ( m_x^4 + m_y^4 +m_z^4).
\end{equation}
By using the identity $m_x^2+m_y^2+m_z^2=1$, it is straightforward to see that the Hamiltonian [Eq.~(\ref{eq_cubic_ham})]
is equivalent to the form 
\begin{equation}
\mathcal{H}_\mathrm{an}^c = -2 K_c \sum_i ( m_x^2 m_y^2 + m_y^2 m_z^2 +m_z^2 m_x^2).
\end{equation}

\subsection{Zeeman energy}
\label{core_eqs:zeeman-energy}
The effective field of a magnetic moment in the presence of an external field $\mvec{H}_z$ is $\mvec{H}_z$ itself, which also can be seen from the Zeeman energy defined as
\begin{gather}
\mathcal{H}_{z} = -  \mu_s \sum_{i}  \mvec{m}_i \cdot \mvec{H}_z.
\end{gather}

\section{LLG equation}
The dynamics of magnetization or magnetic moments is governed by the Landau-Lifshitz-Gilbert (LLG) equation, which can be derived 
using the Lagrangian formulation with a Rayleigh's dissipation function~\cite{Gilbert2004}:
\begin{align}\label{eq_llg}
  \frac{\partial\mathbf{m}}{\partial t}=&-\gamma\mathbf{m}\times\mathbf{H}_\mathrm{eff} + \alpha \mathbf{m}\times \frac{\partial\mathbf{m}}{\partial t}
\end{align}
The LLG equation (\ref{eq_llg}) can be cast into the Landau-Lifshitz (LL) form
\begin{equation}\label{eq_llg2}
\frac{\partial \mathbf{m}}{\partial t} = - \frac{\gamma}{1+\alpha^2}  \mathbf{m} \times \Heff -  \frac{\alpha \gamma}{1+\alpha^2} \mathbf{m} \times \left( \mathbf{m} \times \Heff \right).
\end{equation} 
In practice, we need to use the LL form of the LLG equation [Eq.~(\ref{eq_llg2})] for numerical implementation.
In MicroMagnetic.jl, to solve the LLG equation, we have implemented the adaptive Dormand-Prince method, a member of 
the Runge–Kutta family of ODE solvers~\cite{Press1992}. Note that the typical Runge-Kutta method does 
not conserve the magnitude of magnetization. Therefore, a projection onto a unit sphere
is taken after each successful step to enforce the constraint $|\mathbf{m}|=1$.

\subsection{Spin transfer torques}
The discovery of giant magnetoresistance (GMR) shows that the resistance of a ferromagnetic conductor
depends on its magnetization configuration. This suggests an interaction between the conduction 
electrons and the magnetization, which changes its electric conductivity. In the opposite direction, 
the flow of an electric current will affect the magnetization dynamics. Indeed, conduction electrons
transfer the spin angular momentum to the magnetization of a ferromagnet~\cite{Slonczewski1996}.
Interestingly, Zhang and Li found that~\cite{Zhang2004} most of the physics on the interplay between the magnetization dynamics 
of local moments and the spin-polarized transport of itinerant electrons can be captured mainly by the s-d model.  
The extended LLG equation with spin transfer torques (Zhang-Li model) is given by~\cite{Zhang2004},
\begin{align}\label{eq_llg_stt}
  \frac{\partial\mathbf{m}}{\partial t}=&-\gamma\mathbf{m}\times\mathbf{H}_\mathrm{eff} + \alpha \mathbf{m}\times \frac{\partial\mathbf{m}}{\partial t} \nonumber
  \\ &-\left(\mathbf{u}\cdot\mathbf{\nabla}\right)\mathbf{m}+\beta[\mathbf{m}\times\left(\mathbf{u}\cdot\mathbf{\nabla}\right)\mathbf{m}].
\end{align}
Here, the parameter $\mathbf{u}$ is defined as
\begin{equation}\label{eq_2}
 \mathbf{u}=-\frac{gP\mu_B}{2e M_s}\mathbf{j} = -\frac{P g \mu_B a^3}{2 e \mu_s} \mathbf{j}, 
\end{equation}
where $g$ is the Landé factor, $\mu_{B}$ denotes the Bohr magneton, $e (>0)$ represents the electron charge, 
$M_s$ stands for the saturation magnetization, $P$ is the polarization rate of the current, 
and $\mathbf{j}$ denotes the current density. The $\beta$ term indicates the strength of nonadiabatic spin transfer torques, 
which is very important for the domain-wall motion.

In multilayers, the current flows perpendicular to the plane, which corresponds to the Slonczewski torque:
\begin{align}\label{eq_llg_stt2}
  \frac{\partial \mathbf{m}}{\partial t} =& - \gamma \mathbf{m} \times \mathbf{H} + \alpha \mathbf{m} \times  \frac{\partial \mathbf{m}}{\partial t} \nonumber 
  \\& - a_J \mathbf{m} \times (\mathbf{m} \times \mathbf{p})
 -  b_J \mathbf{m} \times \mathbf{p}
\end{align}
where $\mathbf{p}$ is a unit vector indicating the current polarization direction. 

Although the physical origins of spin-orbit torque (SOT) are entirely different from spin transfer torque, 
The equation can still describe the mathematical form of SOT (\ref{eq_llg_stt2}).
Therefore, in MicroMagnetic.jl, we have implemented the equation (\ref{eq_llg_stt2}), where the user needs to provide parameters $a_J$ 
and $b_J$ to describe different physical processes.

Moreover, note that $-(\mathbf{u} \cdot \mathbf{\nabla})\mathbf{m} = \mathbf{m} \times [\mathbf{m} \times (\mathbf{u} \cdot \mathbf{\nabla})\mathbf{m}]$, 
so all the torques in the extended LLG equations (\ref{eq_llg_stt}) and (\ref{eq_llg_stt2}) begin with $\mathbf{m} \times$. Therefore, these torques 
can be incorporated into the effective fields \(\mathbf{H}_{\text{eff}}\), which also simplifies the code structure~\cite{Meo2023}.

\subsection{Thermal effects}
When the spin system is connected to a thermal reservoir, the effect of temperature can be modeled by adding a stochastic field $\bm{\xi}$
to the effective field $\Heff$ in the standard LLG equation [\ref{eq_llg}], forming the stochastic LLG (SLLG) equation.
The thermal fluctuation is assumed to be a Gaussian white noise, i.e.,
the thermal noise $\bm{\xi}$ obeys the properties
\begin{gather}
\begin{split}\left< \bm{\xi} \right> = 0, \;\;\; \left< \bm{\xi}_i^u \cdot \bm{\xi}_j^v \right> = 2 D \delta_{ij} \delta_{uv},\end{split}
\end{gather}
where $i$ and $j$ are Cartesian indices, $u$ and $v$ indicate the magnetization components and $\left< \cdot \right>$
represents the average taken over different realizations of the fluctuating field.
The constant $D = \alpha k_B T/(\gamma \mu_s)$ denotes the strength of the thermal fluctuations. For the micromagnetic case, $D = \alpha k_B T/(\mu_0 M_s \gamma V_i)$ 
with $V_i$ the volume of the cell.
The SLLG equation can be solved using numerical schemes~\cite{Wang2015}. Alternatively, an effective field can be incorporated into existing integrators:
\begin{gather}
\bm{\xi}^u_i = \eta \sqrt \frac{2 \alpha k_B T}{\mu_0 M_s \gamma V_i dt}
\end{gather}
where $\eta$ is a random number following a normal distribution with a mean of 0 and a standard deviation of 1.

\subsection{Cayley transform}
The dynamics of magnetization evolve on a sphere $S^2$. Thus, there exists a $3\times 3$ rotation matrix $\mathbf{Q}$ such that~\cite{Krishnaprasad2001a}
\begin{equation}
  \mvec{m}(t) =  \mathbf{Q}(t) \mvec{m}(0)
\end{equation}
where $ \mathbf{Q}(0)=\mathbf{I}$ is the identity matrix. 
Notice that the LLG equation (including the spin transfer torques) can be rewritten as
\begin{equation}
  \frac{\partial \mvec{m}}{\partial t} = \mathbf{f}(\mvec{m}) \times \mvec{m} =  \operatorname{skew}[\mathbf{f}] \mvec{m} 
\end{equation}
where $\mathbf{f}= \gamma (\Heff  + \alpha \mvec{m} \times \Heff)/(1+\alpha^2)$ for the LLG equation and the 
contribution of spin transfer torques is $\mathbf{f}_\mathrm{stt} = [(\beta-\alpha)\boldsymbol{\tau} + (1+\alpha\beta) \mvec{m} \times \boldsymbol{\tau})]/(1+\alpha^2) $
with $\boldsymbol{\tau} = (\mvec{u} \cdot \nabla) \mvec{m}$. The $\operatorname{skew}$ operator is defined as 
\begin{equation}
  \operatorname{skew}\left[\boldsymbol{x}\right]=\left(\begin{array}{ccc}
  0 & -x_3 & x_2 \\
  x_3 & 0 & -x_1 \\
  -x_2 & x_1 & 0
  \end{array}\right).
\end{equation}
Therefore, the LLG equation becomes 
\begin{equation}\label{eq_Qt}
  \frac{\partial \mvec{Q}}{\partial t} = \operatorname{skew}[\mathbf{f}] \mvec{Q}(t).
\end{equation}
The equation (\ref{eq_Qt}) can be further reformed using the Cayley transform, which is defined as
\begin{equation}\label{eq_cay}
  \mvec{Q} =  \left(\mathbf{I} - \tfrac{1}{2} \mathbf{\Omega} \right)^{-1} \left(\mathbf{I} + \tfrac{1}{2} \mathbf{\Omega} \right)
\end{equation}
where $\mathbf{\Omega} = \operatorname{skew}[\boldsymbol{\omega}]$.  The factor of $1/2$ in the Cayley transform
can be changed to other nonzero numbers.  Taking the derivative concerning time on both sides of the equation (\ref{eq_cay}) 
yields $\frac{1}{2} \dot{\mvec{\Omega}} (\mvec{Q}+\mathbf{I})= (\mvec{I} - \frac{1}{2} \mvec{\Omega}) \dot{\mvec{Q}}$.
Noticed $\mvec{Q}+\mathbf{I} = 2 (\mathbf{I} - \frac{1}{2} \mathbf{\Omega})^{-1}$, 
one obtains the dcayinv equation~\cite{Iserles2000}
\begin{equation}\label{eq_dcay}
  \frac{\partial \mathbf{\Omega}}{\partial t} = \mathbf{F} - \frac{1}{2} [\mathbf{\Omega}, \mathbf{F}]
  - \frac{1}{4} \mathbf{\Omega} \mathbf{F} \mathbf{\Omega},
\end{equation}
where $\mathbf{F} = \operatorname{skew}[\mathbf{f}]$. 
In MicroMagnetic.jl, we make use of the vector form of equation (\ref{eq_dcay}), which reads
\begin{equation}\label{eq_omega}
  \frac{\partial \boldsymbol{\omega}}{\partial t} = \mathbf{f} - \frac{1}{2}
  (\boldsymbol{\omega} \times \mathbf{f} )
  + \frac{1}{4} (\boldsymbol{\omega} \cdot \mathbf{f}) \boldsymbol{\omega}.
\end{equation}
The equation (\ref{eq_omega}) can be integrated using the normal integration algorithms such as Runge-Kutta
without worrying the conservation property~\cite{Krishnaprasad2001a}.
It's worth mentioning that the absolute value of \(\omega\) will always increase for a nonzero \(\mathbf{f}\). Therefore, 
restarting procedures are necessary to prevent \(|\omega|\) from becoming very large~\cite{Diele1998a}.

\section{Steepest descent energy minimization}
A prevalent task in micromagnetics is finding the local energy minima states, for instance, calculating the hysteresis loops.
The energy minimization process can be achieved by directly solving the LLG equation with a positive Gilbert damping.
However, the steepest descent methods combined with step-sized selected using Barzilian-Borwein rules have been proven a much more efficient way
to find the local energy minima~\cite{Abert2014a, Exl2014a}. 
The MicroMagnetic.jl implementation of the steepest descent method uses the following update, 
\begin{equation}\label{eq_update}
  \mvec{m}_{k+1}=\mvec{m}_{k}-{\tau}_k \frac{\mvec{m}_{k}+\mvec{m}_{k+1}}{2} \times\left[\mvec{m}_{k} \times \Heff (\mvec{m}_{k})\right]
\end{equation}
which preserves the modulus of $\mvec{m}$, i.e., $\mvec{m}_{k+1}^2 = \mvec{m}_{k}^2$. The equation (\ref{eq_update}) explicitly gives 
\begin{equation}
  (1+\frac{{\tau}_k^2}{4} \mvec{f}_k^2)\mathbf{m}_{k+1} =
  (1-\frac{{\tau}_k^2}{4} \mvec{f}_k^2)\mathbf{m}_{k} -  {\tau}_k \mathbf{g}_k
\end{equation}
where $\mathbf{f}_k  = \mathbf{m}_k \times \Heff$ and $\mvec{g}_{k} =\mvec{m}_{k} \times\left(\mvec{m}_{k} \times \Heff \right)$.
The step size $\tau_k$ can be chosen using Barzilian-Borwein rules:
\begin{equation}
\tau_{k}^{1}=\frac{\sum_{i} \mvec{s}_{k-1}^{i} \cdot \mvec{s}_{k-1}^{i}}{\sum_{i} \mvec{s}_{k-1}^{i} \cdot \mvec{y}_{k-1}^{i}}, \quad 
\quad \tau_{k}^{2}=\frac{\sum_{i} \mvec{s}_{k-1}^{i} \cdot \mvec{y}_{k-1}^{i}}{\sum_{i} \mvec{y}_{k-1}^{i} \cdot \mvec{y}_{k-1}^{i}}
\end{equation}
where $\mvec{s}_{k-1} =\mvec{m}_{k}-\mvec{m}_{k-1}$ and $\mvec{y}_{k-1} =\mvec{g}_{k}-\mvec{g}_{k-1}$.

\section{Miscellaneous}
In MicroMagnetic.jl, we provided a Markov chain Monte Carlo implementation for the atomistic spin model. The implemented Hamiltonian includes
exchange interaction $\mathcal{H}_\mathrm{ex}$, anisotropy interaction $\mathcal{H}_\mathrm{an}$, 
Dzyaloshinskii-Moriya interaction $\mathcal{H}_\mathrm{dmi}$, and Zeeman interaction $\mathcal{H}_z$.
The parallel scheme for the Metropolis algorithm relies on the relative positions within the lattice. 
For example, in a cubic lattice when considering only nearest neighbor interactions, we can divide sites into three classes based on the remainder of their positions (i+j+k) modulo 3, 
and each class can be run in parallel.

The nudged elastic band (NEB) method is a chain method to compute the minimum energy path (MEP) between two states. Initially, a chain comprising multiple images, 
each representing a copy of the magnetization, is constructed, and then the whole system is relaxed. The two end images corresponding 
to the initial and final states are fixed, representing the energy states specified by the user. The system, encompassing all free images, 
undergoes relaxation to minimize the total energy, akin to relaxing a magnetic system using the LLG equation with the precession term deactivated. 
A notable distinction lies in the effective field: in the LLG equation, it stems from the functional derivative of the system energy concerning magnetization, 
whereas in NEB, the effective field of each image additionally incorporates the influence of its adjacent neighbors (i.e., images $n-1$ and $n+1$).
The so-called tangents describe this influence: only the perpendicular part of the effective field is retained when relaxing the entire system~\cite{Cortes2017}.

\section{Verifications and Performance}
\subsection{A magnetic moment under an external magnetic field}
\begin{figure}[htbp]
\begin{center}
\includegraphics[width=0.4\textwidth]{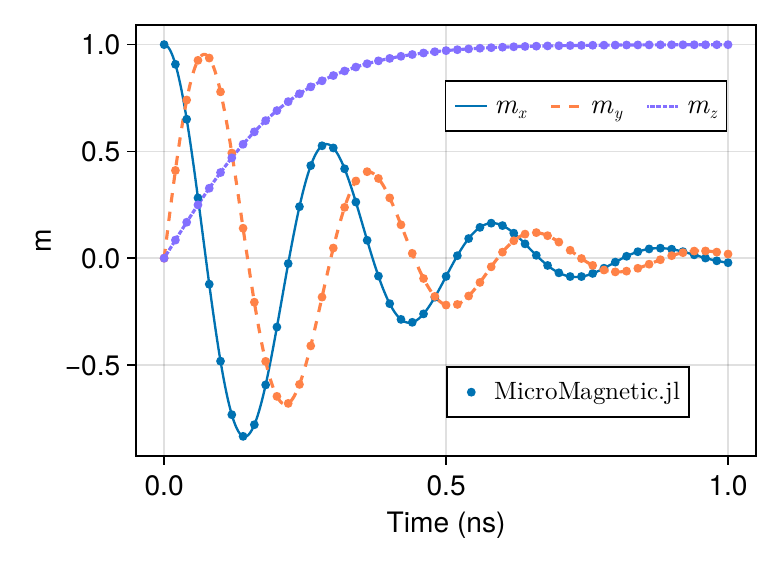} 
\caption{The precession motion of a magnetic moment under an external field with $H_z=1\times10^{5}$ A/m.  
The parameters used are damping constant $\alpha=0.5$ and gyromagnetic ratio $\gamma=2.21 \times 10^5$ m/(A$\cdot$s).}
\label{fig_single_spin}
\end{center}
\end{figure}

The LLG equation is nonlinear. The simplest case is the precession motion of a magnetic moment under an external magnetic field. 
Assuming the external field is along with the $z$-axis and the initial state is $\mathbf{m}_0=(1,0,0)$, then we have
\begin{gather}\label{eq_llg_m}
\begin{split}
m_x(t) &= \cos(\tilde{\gamma} H_z t)/\cosh(\alpha \tilde{\gamma} H_z t), \\
m_y(t) &= \sin(\tilde{\gamma} H_z t)/\cosh(\alpha \tilde{\gamma} H_z t), \\
m_z(t) &= \tanh(\alpha \tilde{\gamma} H_z t), \\
\end{split}
\end{gather}
where $\tilde{\gamma}=\gamma/(1+\alpha^2)$. Figure~\ref{fig_single_spin} shows the precessional motion of the macrospin under 
an external magnetic field where the (dashed) lines represent the analytical prediction [Eq.~(\ref{eq_llg_m})]. 
The macrospin will align with the external field after dissipating its energy.

\begin{figure}[htbp]
  \begin{center}
  \includegraphics[width=0.4\textwidth]{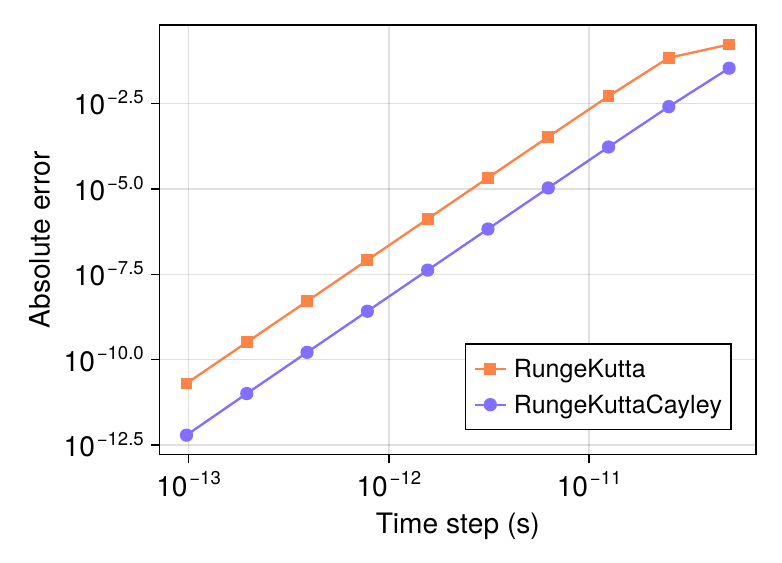} 
  \caption{The absolute error of the precession motion of a magnetic moment under an external field at $t=1$ ns. 
  A small damping $\alpha=0.05$ is used.}
  \label{fig_error}
  \end{center}
\end{figure}

The analytical results provided by Eq.~(\ref{eq_llg_m}) enable us to evaluate the performance of different integration algorithms. 
We compared the classical Runge-Kutta method with projection and one utilizing a Cayley transformation. The absolute error of the precessional motion 
at $t=1$ ns for both approaches is depicted in Figure \ref{fig_error}, with the absolute error defined as $|\mvec{m}_\mathrm{an}-\mvec{m}|$.
The Cayley transform version performs much better than the projection method for the classical Runge-Kutta scheme.

\subsection{Domain-wall motion under currents}
A few exact spatiotemporal solutions for the LLG equation have been reported in the literature. One example is a 1D
head-to-head domain wall driven by STT under an arbitrary time ­dependence function $u=u(t)$~\cite{Wang2015, Wang2018}
\begin{equation}\label{eq_dw_v}
  \dot{\phi}=\frac{(\alpha-\beta)u(t)}{(1+\alpha^2)\Delta}, \qquad \dot{z}_*=-\frac{(1+\alpha\beta)u(t)}{(1+\alpha^2)} .
\end{equation}
where $\dot{z}_*$ is the domain wall center, $\phi$ is the domain wall tilt angle and $\Delta = \sqrt{A/K}$ is the width of domain wall.

\begin{figure}[htbp]
\begin{center}
\includegraphics[width=0.4\textwidth]{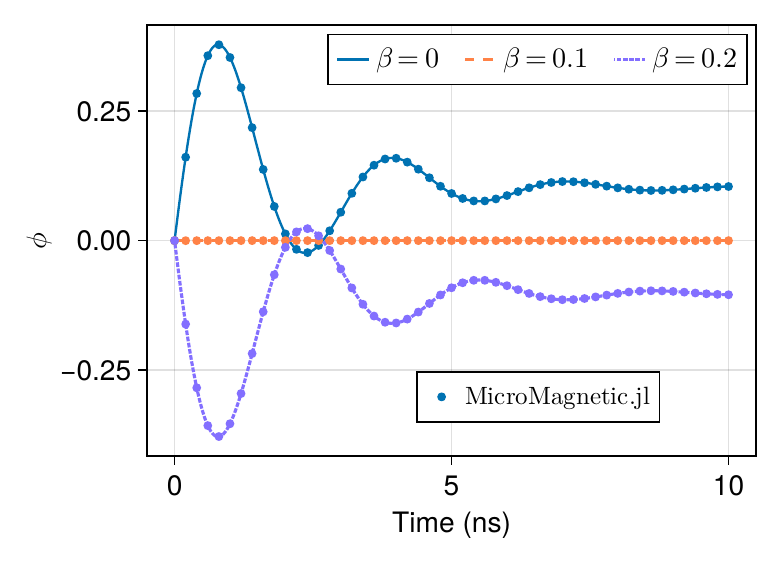} 
\caption{A comparison between the simulation and the analytical solution [Eq.~(\ref{eq_dw_v})] for the motion of a head-to-head domain wall.}
\label{fig_dw_stt}
\end{center}
\end{figure}

To validate MicroMagnetic.jl against the analytical solution [Eq.~(\ref{eq_dw_v})], we apply a time-dependent 
current represented as $u(t)=u_0 e^{-b t} \cos(\omega  t)$ to the system in our simulation, 
where $b=5 \times 10^8$ 1/s and $\omega =2 \times 10^{9}$ 1/s are employed.
We conduct simulations with the parameters specific to Permalloy: the saturation magnetization $M_s = 8\times 10^5\,\mathrm{A/m}$, 
the exchange constant $A = 1.3 \times 10^{-11}\, \mathrm{J/m}$, and an effective anisotropy $K=1\times 10^5\,\mathrm{J/m^3}$. 
We maintain a fixed damping constant $\alpha=0.1$ throughout the simulations while varying $\beta$. 
As depicted in Figure~\ref{fig_dw_stt}, the comparison between the domain-wall tilt angle $\phi$ obtained using MicroMagnetic.jl and 
the analytical equation (\ref{eq_dw_v}) demonstrates a strong agreement.

\subsection{Equilibrium distribution of a nanoparticle}
\begin{figure}[tbhp]
\begin{center}
\includegraphics[width=0.4\textwidth]{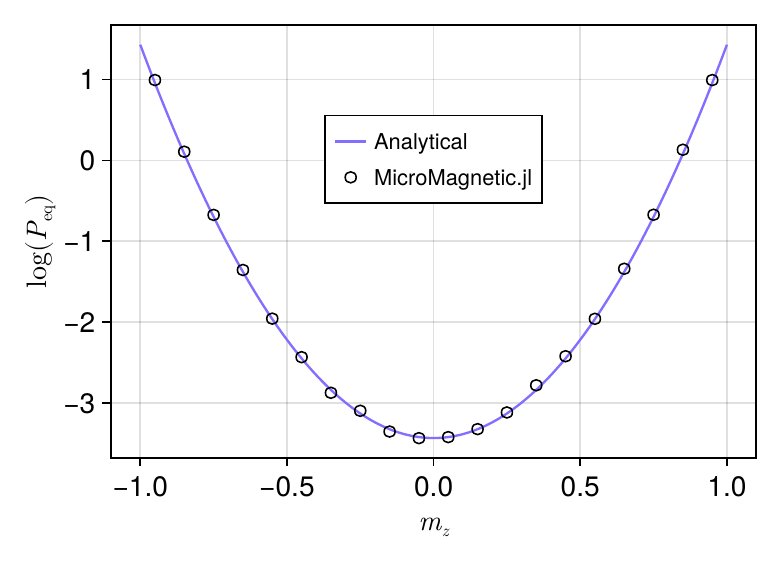} 
\caption{The equilibrium probability distribution  $P_\mathrm{eq}(m_z)$ of a nanoparticle at $T=300$ K.}
\label{fig_sllg_pb}
\end{center}
\end{figure}
As shown in Ref.~\cite{Wang2015}, we consider a nanoparticle with an effective uniaxial anisotropy along $z$-direction and its energy density reads
\begin{equation}
u =  K(1-m_z^2).
\end{equation} 
In the equilibrium state, the probability distribution function (PDF) $P_\mathrm{eq}(m_z)$ satisfies $P_\mathrm{eq}(m_z) \propto \exp[- u V/(k_B T) ]$.
Hence, we have
\begin{equation}
P_\mathrm{eq}(m_z) = \frac{1}{Z} e^{- \chi (1- m_z^2)},
\end{equation}
where $\chi =KV/(k_B T)$ and $Z=\int e^{- \chi (1- m_z^2)} \,dm_z$ is the partition function. 
Assume the volume $V$ of the particle is $V=2.8\times10^{-26}$ nm$^3$, $M_s=1.42\times 10^{6}$ A/m and $K= 7.2\times 10^5 $J/m$^3$. 
We obtain $Z=0.428$ for $T=300$ K. 
Figure~\ref{fig_sllg_pb} shows the equilibrium probability distribution $P_\mathrm{eq}(m_z)$ of a nanoparticle at $T=300$ K, 
where $\alpha=0.1$ are used in the simulation.

\subsection{Monte Carlo simulation}
\begin{figure}[htbp]
  \begin{center}
  \includegraphics[width=0.45\textwidth]{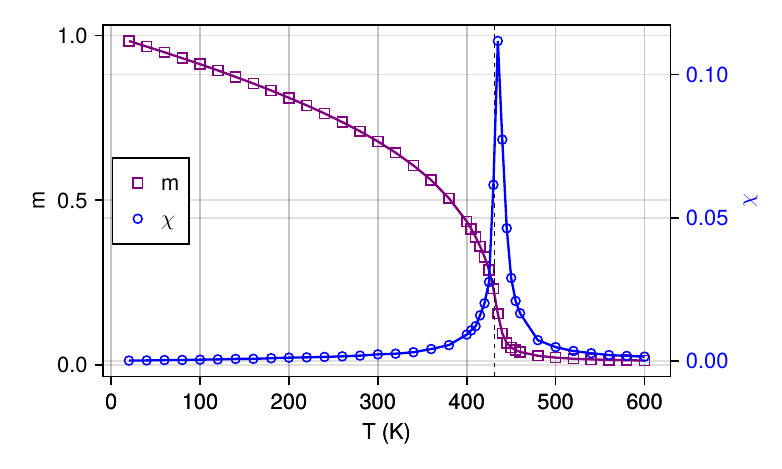} 
  \caption{The simulated temperature-dependent magnetization and susceptibility using Monte Carlo methods.}
  \label{fig_mc}
  \end{center}
\end{figure}
Besides directly solving the stochastic LLG equation, in MicroMagnetic.jl, we can also use Monte Carlo to compute the M-T curve.
For the atomistic model with $z$ nearest neighbors, the relation between exchange constant and $T_c$ reads
\begin{equation} \label{eq_Tc}
J = \frac{3 k_B T_c}{ \epsilon z }
\end{equation}
where $\epsilon$ is a correction factor and for 3D classical Heisenberg model $\epsilon \approx 0.719$~\cite{Garanin1996}. 
Figure~\ref{fig_mc} presents the calculated average magnetization $m$ and magnetic susceptibility $\chi$ as functions of temperature, 
using the Monte Carlo method with a cubic mesh of size $N = 30 \times 30 \times 30$. The magnetic susceptibility is computed as
\begin{equation} \label{eq_chi}
  \chi = \frac{N}{T}\left [\langle m^2 \rangle - \langle m \rangle^2 \right ].
\end{equation}
The critical temperature $T_c \sim 431$ K, determined for $J = 300 \, k_B$ using equation (\ref{eq_Tc}), is indicated by the dashed vertical line in Fig.~\ref{fig_mc}. 
It can be observed that the temperature corresponding to the maximum value of $\chi$ closely matches the theoretical prediction.

\subsection{Standard Problem 4}
\begin{figure}[htbp]
  \begin{center}
  \includegraphics[width=0.4\textwidth]{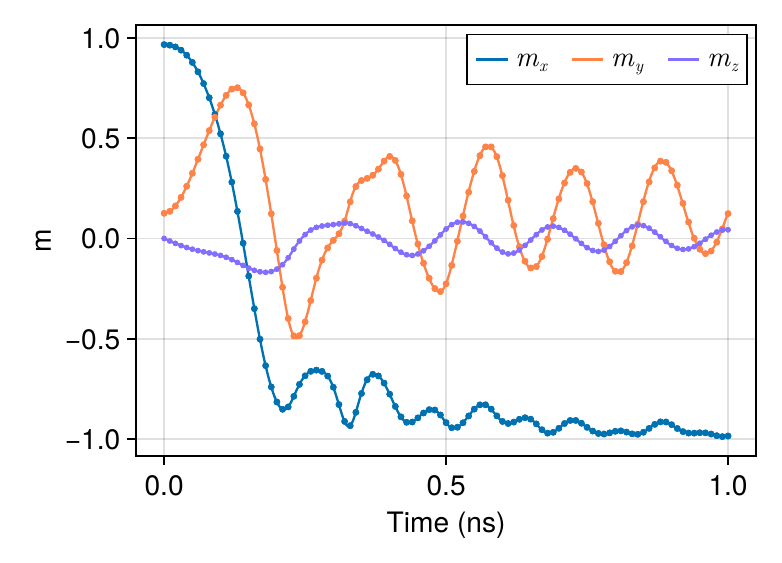} 
  \caption{The calculation of the standard problem 4 using OOMMF (lines) and MicroMagnetic.jl (dots).}
  \label{fig_std4}
  \end{center}
\end{figure}

Micromagnetic Standard Problem 4 is a benchmark designed to validate the accuracy of micromagnetic simulations. 
The system under study is a thin film of Permalloy with dimensions of 500 nm in length, 125 nm in width, and 3 nm in thickness. 
Standard Problem 4 consists of two parts: The first part involves relaxing the system from an initial uniform state $\mathbf{m_0} = (1, 0.25, 0.1)$, 
where the system's energy is iteratively minimized until the change in magnetization falls below a specified threshold, 
ultimately forming an "S" state. After achieving the "S" state, the next step is to apply an external magnetic field $\mathbf{H}=(-24.6\mathrm{mT}, 4.3\mathrm{mT}, 0)$
and observe the resulting magnetization dynamics. Figure \ref{fig_std4} shows the evolution of the average magnetization as a 
function of time after applying the external magnetic field, using both OOMMF and MicroMagnetic.jl. 
The results demonstrate a strong agreement between the calculated magnetizations from both methods.

\subsection{Performance}
\begin{figure}[htbp]
  \begin{center}
  \includegraphics[width=0.4\textwidth]{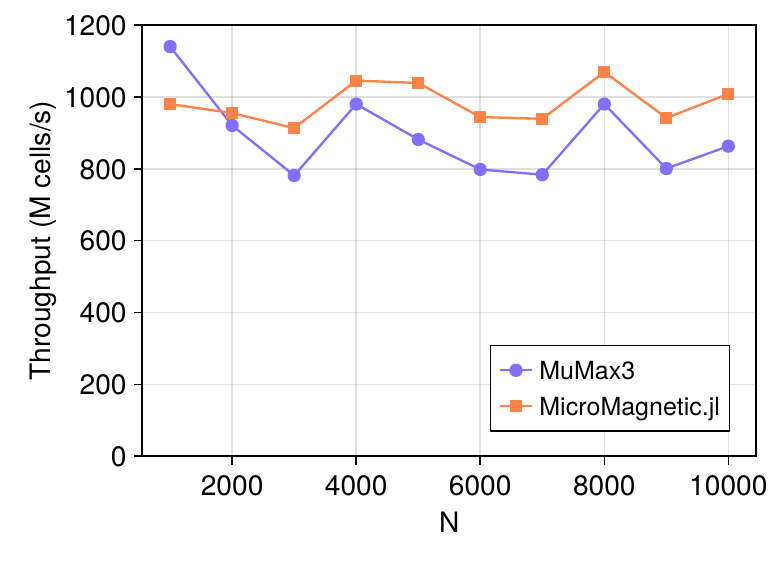} 
  \caption{The throughput of MuMax3 and MicroMagnetic.jl on an NVIDIA GPU (A800 SXM4 80GB). The system size is $N \times N \times 1$, and demagnetization, exchange, and Zeeman interactions are considered. The Henu integrators are used for testing.}
  \label{fig_th}
  \end{center}
\end{figure}

In micromagnetic simulations, the most time-consuming component is typically the computation of the demagnetizing field. 
Despite being accelerated by FFT, this calculation still dominates the total simulation time. To evaluate the performance of 
MicroMagnetic.jl, we compared it with MuMax3. Figure~\ref{fig_th} presents the performance comparison between the two, 
using an \(N \times N \times 1\) system. The results indicate that on the NVIDIA A800 GPU, both MuMax3 and MicroMagnetic.jl 
achieve throughput rates of up to 1000 million cells per second, with MicroMagnetic.jl performing similarly to MuMax3.

\section{Program availability}
MicroMagnetic.jl is an open-source Julia package licensed under the MIT license, and it can be installed directly through Julia's package manager, 
\texttt{Pkg}. The source code is available at https://github.com/ww1g11/MicroMagnetic.jl.

\section{summary}
In conclusion, MicroMagnetic.jl has been introduced as a Julia package designed for micromagnetic and atomistic simulations. 
With support for CPU and various GPU platforms, researchers can efficiently explore magnetic structures and phenomena 
at the micrometer scale. MicroMagnetic.jl's open-source nature and ease of extensibility are intended to foster collaboration 
and innovation, inviting developers to contribute code and expand its functionality. MicroMagnetic.jl's design maximizes user flexibility, 
making it a versatile toolset for researchers seeking advancements in the study of magnetic materials and devices.

\section{acknowledgments}
We acknowledge financial support from the National Key R\&D Program of China (Grant No. 2022YFA1403603) and 
the Strategic Priority Research Program of Chinese Academy of Sciences (Grant No. XDB33030100), 
the National Natural Science Funds for Distinguished Young Scholar (52325105),
the National Natural Science Foundation of China (Grants No. 12374098, No. 11974021, and No. 12241406)
and the CAS Project for Young Scientists in Basic Research (YSBR-084). 

\bibliographystyle{apsrev4-2}
\bibliography{micro}

\begin{thebibliography}{48}%
\makeatletter
\providecommand \@ifxundefined [1]{%
 \@ifx{#1\undefined}
}%
\providecommand \@ifnum [1]{%
 \ifnum #1\expandafter \@firstoftwo
 \else \expandafter \@secondoftwo
 \fi
}%
\providecommand \@ifx [1]{%
 \ifx #1\expandafter \@firstoftwo
 \else \expandafter \@secondoftwo
 \fi
}%
\providecommand \natexlab [1]{#1}%
\providecommand \enquote  [1]{``#1''}%
\providecommand \bibnamefont  [1]{#1}%
\providecommand \bibfnamefont [1]{#1}%
\providecommand \citenamefont [1]{#1}%
\providecommand \href@noop [0]{\@secondoftwo}%
\providecommand \href [0]{\begingroup \@sanitize@url \@href}%
\providecommand \@href[1]{\@@startlink{#1}\@@href}%
\providecommand \@@href[1]{\endgroup#1\@@endlink}%
\providecommand \@sanitize@url [0]{\catcode `\\12\catcode `\$12\catcode
  `\&12\catcode `\#12\catcode `\^12\catcode `\_12\catcode `\%12\relax}%
\providecommand \@@startlink[1]{}%
\providecommand \@@endlink[0]{}%
\providecommand \url  [0]{\begingroup\@sanitize@url \@url }%
\providecommand \@url [1]{\endgroup\@href {#1}{\urlprefix }}%
\providecommand \urlprefix  [0]{URL }%
\providecommand \Eprint [0]{\href }%
\providecommand \doibase [0]{https://doi.org/}%
\providecommand \selectlanguage [0]{\@gobble}%
\providecommand \bibinfo  [0]{\@secondoftwo}%
\providecommand \bibfield  [0]{\@secondoftwo}%
\providecommand \translation [1]{[#1]}%
\providecommand \BibitemOpen [0]{}%
\providecommand \bibitemStop [0]{}%
\providecommand \bibitemNoStop [0]{.\EOS\space}%
\providecommand \EOS [0]{\spacefactor3000\relax}%
\providecommand \BibitemShut  [1]{\csname bibitem#1\endcsname}%
\let\auto@bib@innerbib\@empty
\bibitem [{\citenamefont {Kronm{\"u}ller}(2007)}]{Kronmuller2007}%
  \BibitemOpen
  \bibfield  {author} {\bibinfo {author} {\bibfnamefont {H.}~\bibnamefont
  {Kronm{\"u}ller}},\ }in\ \href {https://doi.org/10.1002/9780470022184.hmm201}
  {\emph {\bibinfo {booktitle} {Handbook of {{Magnetism}} and {{Advanced
  Magnetic Materials}}}}},\ \bibinfo {editor} {edited by\ \bibinfo {editor}
  {\bibfnamefont {H.}~\bibnamefont {Kronm{\"u}ller}}\ and\ \bibinfo {editor}
  {\bibfnamefont {S.}~\bibnamefont {Parkin}}}\ (\bibinfo  {publisher} {John
  Wiley \& Sons, Ltd},\ \bibinfo {address} {Chichester, UK},\ \bibinfo {year}
  {2007})\BibitemShut {NoStop}%
\bibitem [{\citenamefont {Leliaert}\ and\ \citenamefont
  {Mulkers}(2019)}]{Leliaert2019}%
  \BibitemOpen
  \bibfield  {author} {\bibinfo {author} {\bibfnamefont {J.}~\bibnamefont
  {Leliaert}}\ and\ \bibinfo {author} {\bibfnamefont {J.}~\bibnamefont
  {Mulkers}},\ }\href {https://doi.org/10.1063/1.5093730} {\bibfield  {journal}
  {\bibinfo  {journal} {Journal of Applied Physics}\ }\textbf {\bibinfo
  {volume} {125}},\ \bibinfo {pages} {180901} (\bibinfo {year}
  {2019})}\BibitemShut {NoStop}%
\bibitem [{\citenamefont {Rezende}(2020)}]{Rezende2020}%
  \BibitemOpen
  \bibfield  {author} {\bibinfo {author} {\bibfnamefont {S.~M.}\ \bibnamefont
  {Rezende}},\ }\href {https://doi.org/10.1007/978-3-030-41317-0} {\emph
  {\bibinfo {title} {Fundamentals of {{Magnonics}}}}},\ \bibinfo {series}
  {Lecture {{Notes}} in {{Physics}}}, Vol.\ \bibinfo {volume} {969}\ (\bibinfo
  {publisher} {Springer International Publishing},\ \bibinfo {address} {Cham},\
  \bibinfo {year} {2020})\BibitemShut {NoStop}%
\bibitem [{\citenamefont {Flebus}\ \emph {et~al.}(2024)\citenamefont {Flebus},
  \citenamefont {Grundler}, \citenamefont {Rana}, \citenamefont {Otani},
  \citenamefont {Barsukov}, \citenamefont {Barman}, \citenamefont {Gubbiotti},
  \citenamefont {Landeros}, \citenamefont {Akerman}, \citenamefont {Ebels},
  \citenamefont {Pirro}, \citenamefont {Demidov}, \citenamefont {Schultheiss},
  \citenamefont {Csaba}, \citenamefont {Wang}, \citenamefont {Nikonov},
  \citenamefont {Ciubotaru}, \citenamefont {Che}, \citenamefont {Hertel},
  \citenamefont {Ono}, \citenamefont {Afanasiev}, \citenamefont {Mentink},
  \citenamefont {Rasing}, \citenamefont {Hillebrands}, \citenamefont
  {Viola~Kusminskiy}, \citenamefont {Zhang}, \citenamefont {Du}, \citenamefont
  {Finco}, \citenamefont {Van Der~Sar}, \citenamefont {Luo}, \citenamefont
  {Shiota}, \citenamefont {Sklenar}, \citenamefont {Yu},\ and\ \citenamefont
  {Rao}}]{Flebus2024}%
  \BibitemOpen
  \bibfield  {author} {\bibinfo {author} {\bibfnamefont {B.}~\bibnamefont
  {Flebus}}, \bibinfo {author} {\bibfnamefont {D.}~\bibnamefont {Grundler}},
  \bibinfo {author} {\bibfnamefont {B.}~\bibnamefont {Rana}}, \bibinfo {author}
  {\bibfnamefont {Y.}~\bibnamefont {Otani}}, \bibinfo {author} {\bibfnamefont
  {I.}~\bibnamefont {Barsukov}}, \bibinfo {author} {\bibfnamefont
  {A.}~\bibnamefont {Barman}}, \bibinfo {author} {\bibfnamefont
  {G.}~\bibnamefont {Gubbiotti}}, \bibinfo {author} {\bibfnamefont
  {P.}~\bibnamefont {Landeros}}, \bibinfo {author} {\bibfnamefont
  {J.}~\bibnamefont {Akerman}}, \bibinfo {author} {\bibfnamefont {U.~S.}\
  \bibnamefont {Ebels}}, \bibinfo {author} {\bibfnamefont {P.}~\bibnamefont
  {Pirro}}, \bibinfo {author} {\bibfnamefont {V.~E.}\ \bibnamefont {Demidov}},
  \bibinfo {author} {\bibfnamefont {K.}~\bibnamefont {Schultheiss}}, \bibinfo
  {author} {\bibfnamefont {G.}~\bibnamefont {Csaba}}, \bibinfo {author}
  {\bibfnamefont {Q.}~\bibnamefont {Wang}}, \bibinfo {author} {\bibfnamefont
  {D.~E.}\ \bibnamefont {Nikonov}}, \bibinfo {author} {\bibfnamefont
  {F.}~\bibnamefont {Ciubotaru}}, \bibinfo {author} {\bibfnamefont
  {P.}~\bibnamefont {Che}}, \bibinfo {author} {\bibfnamefont {R.}~\bibnamefont
  {Hertel}}, \bibinfo {author} {\bibfnamefont {T.}~\bibnamefont {Ono}},
  \bibinfo {author} {\bibfnamefont {D.}~\bibnamefont {Afanasiev}}, \bibinfo
  {author} {\bibfnamefont {J.~H.}\ \bibnamefont {Mentink}}, \bibinfo {author}
  {\bibfnamefont {T.}~\bibnamefont {Rasing}}, \bibinfo {author} {\bibfnamefont
  {B.}~\bibnamefont {Hillebrands}}, \bibinfo {author} {\bibfnamefont
  {S.}~\bibnamefont {Viola~Kusminskiy}}, \bibinfo {author} {\bibfnamefont
  {W.}~\bibnamefont {Zhang}}, \bibinfo {author} {\bibfnamefont {C.~R.}\
  \bibnamefont {Du}}, \bibinfo {author} {\bibfnamefont {A.}~\bibnamefont
  {Finco}}, \bibinfo {author} {\bibfnamefont {T.}~\bibnamefont {Van Der~Sar}},
  \bibinfo {author} {\bibfnamefont {Y.~K.}\ \bibnamefont {Luo}}, \bibinfo
  {author} {\bibfnamefont {Y.}~\bibnamefont {Shiota}}, \bibinfo {author}
  {\bibfnamefont {J.}~\bibnamefont {Sklenar}}, \bibinfo {author} {\bibfnamefont
  {T.}~\bibnamefont {Yu}},\ and\ \bibinfo {author} {\bibfnamefont
  {J.}~\bibnamefont {Rao}},\ }\bibfield  {journal} {\bibinfo  {journal}
  {Journal of Physics: Condensed Matter}\ }\href
  {https://doi.org/10.1088/1361-648X/ad399c} {10.1088/1361-648X/ad399c}
  (\bibinfo {year} {2024})\BibitemShut {NoStop}%
\bibitem [{\citenamefont {Grollier}\ \emph {et~al.}(2020)\citenamefont
  {Grollier}, \citenamefont {Querlioz}, \citenamefont {Camsari}, \citenamefont
  {{Everschor-Sitte}}, \citenamefont {Fukami},\ and\ \citenamefont
  {Stiles}}]{Grollier2020}%
  \BibitemOpen
  \bibfield  {author} {\bibinfo {author} {\bibfnamefont {J.}~\bibnamefont
  {Grollier}}, \bibinfo {author} {\bibfnamefont {D.}~\bibnamefont {Querlioz}},
  \bibinfo {author} {\bibfnamefont {K.~Y.}\ \bibnamefont {Camsari}}, \bibinfo
  {author} {\bibfnamefont {K.}~\bibnamefont {{Everschor-Sitte}}}, \bibinfo
  {author} {\bibfnamefont {S.}~\bibnamefont {Fukami}},\ and\ \bibinfo {author}
  {\bibfnamefont {M.~D.}\ \bibnamefont {Stiles}},\ }\href
  {https://doi.org/10.1038/s41928-019-0360-9} {\bibfield  {journal} {\bibinfo
  {journal} {Nature Electronics}\ }\textbf {\bibinfo {volume} {3}},\ \bibinfo
  {pages} {360} (\bibinfo {year} {2020})}\BibitemShut {NoStop}%
\bibitem [{\citenamefont {Abert}(2019)}]{Abert2019}%
  \BibitemOpen
  \bibfield  {author} {\bibinfo {author} {\bibfnamefont {C.}~\bibnamefont
  {Abert}},\ }\href {https://doi.org/10.1140/epjb/e2019-90599-6} {\bibfield
  {journal} {\bibinfo  {journal} {The European Physical Journal B}\ }\textbf
  {\bibinfo {volume} {92}},\ \bibinfo {pages} {120} (\bibinfo {year}
  {2019})}\BibitemShut {NoStop}%
\bibitem [{\citenamefont {Barla}\ \emph {et~al.}(2021)\citenamefont {Barla},
  \citenamefont {Joshi},\ and\ \citenamefont {Bhat}}]{Barla2021}%
  \BibitemOpen
  \bibfield  {author} {\bibinfo {author} {\bibfnamefont {P.}~\bibnamefont
  {Barla}}, \bibinfo {author} {\bibfnamefont {V.~K.}\ \bibnamefont {Joshi}},\
  and\ \bibinfo {author} {\bibfnamefont {S.}~\bibnamefont {Bhat}},\ }\href
  {https://doi.org/10.1007/s10825-020-01648-6} {\bibfield  {journal} {\bibinfo
  {journal} {Journal of Computational Electronics}\ }\textbf {\bibinfo {volume}
  {20}},\ \bibinfo {pages} {805} (\bibinfo {year} {2021})}\BibitemShut
  {NoStop}%
\bibitem [{\citenamefont {Miltat}\ and\ \citenamefont
  {Donahue}(2007)}]{Miltat2007}%
  \BibitemOpen
  \bibfield  {author} {\bibinfo {author} {\bibfnamefont {J.~E.}\ \bibnamefont
  {Miltat}}\ and\ \bibinfo {author} {\bibfnamefont {M.~J.}\ \bibnamefont
  {Donahue}},\ }in\ \href {https://doi.org/10.1002/9780470022184.hmm202} {\emph
  {\bibinfo {booktitle} {Handbook of {{Magnetism}} and {{Advanced Magnetic
  Materials}}}}},\ \bibinfo {editor} {edited by\ \bibinfo {editor}
  {\bibfnamefont {H.}~\bibnamefont {Kronm{\"u}ller}}\ and\ \bibinfo {editor}
  {\bibfnamefont {S.}~\bibnamefont {Parkin}}}\ (\bibinfo  {publisher} {John
  Wiley \& Sons, Ltd},\ \bibinfo {address} {Chichester, UK},\ \bibinfo {year}
  {2007})\BibitemShut {NoStop}%
\bibitem [{\citenamefont {Schrefl}\ \emph {et~al.}(2007)\citenamefont
  {Schrefl}, \citenamefont {Hrkac},\ and\ \citenamefont {Bance}}]{Schrefl2007}%
  \BibitemOpen
  \bibfield  {author} {\bibinfo {author} {\bibfnamefont {T.}~\bibnamefont
  {Schrefl}}, \bibinfo {author} {\bibfnamefont {G.}~\bibnamefont {Hrkac}},\
  and\ \bibinfo {author} {\bibfnamefont {S.}~\bibnamefont {Bance}},\ }in\
  \href@noop {} {\emph {\bibinfo {booktitle} {Handbook of {{Magnetism}} and
  {{Advanced Magnetic Materials}}.}}},\ Vol.~\bibinfo {volume} {2}\ (\bibinfo
  {year} {2007})\BibitemShut {NoStop}%
\bibitem [{\citenamefont {Donahue}\ and\ \citenamefont
  {Porter}(1999)}]{Porter1999}%
  \BibitemOpen
  \bibfield  {author} {\bibinfo {author} {\bibfnamefont {M.}~\bibnamefont
  {Donahue}}\ and\ \bibinfo {author} {\bibfnamefont {D.}~\bibnamefont
  {Porter}},\ }\href@noop {} {\bibinfo {title} {{{OOMMF User}}'s {{Guide}},
  {{Version}} 1.0}},\ \bibinfo {howpublished} {http://math.nist.gov/oommf/}
  (\bibinfo {year} {1999})\BibitemShut {NoStop}%
\bibitem [{\citenamefont {Vansteenkiste}\ \emph {et~al.}(2014)\citenamefont
  {Vansteenkiste}, \citenamefont {Leliaert}, \citenamefont {Dvornik},
  \citenamefont {{Garcia-Sanchez}},\ and\ \citenamefont
  {Van~Waeyenberge}}]{Vansteenkiste2014}%
  \BibitemOpen
  \bibfield  {author} {\bibinfo {author} {\bibfnamefont {A.}~\bibnamefont
  {Vansteenkiste}}, \bibinfo {author} {\bibfnamefont {J.}~\bibnamefont
  {Leliaert}}, \bibinfo {author} {\bibfnamefont {M.}~\bibnamefont {Dvornik}},
  \bibinfo {author} {\bibfnamefont {F.}~\bibnamefont {{Garcia-Sanchez}}},\ and\
  \bibinfo {author} {\bibfnamefont {B.}~\bibnamefont {Van~Waeyenberge}},\
  }\href {https://doi.org/10.1063/1.4899186} {\bibfield  {journal} {\bibinfo
  {journal} {AIP ADVANCES}\ }\textbf {\bibinfo {volume} {4}},\ \bibinfo {pages}
  {107133} (\bibinfo {year} {2014})},\ \Eprint
  {https://arxiv.org/abs/1406.7635} {arxiv:1406.7635} \BibitemShut {NoStop}%
\bibitem [{\citenamefont {Bisotti}\ \emph {et~al.}(2018)\citenamefont
  {Bisotti}, \citenamefont {{Cort{\'e}s-Ortu{\~n}o}}, \citenamefont {Pepper},
  \citenamefont {Wang}, \citenamefont {Beg}, \citenamefont {Kluyver},\ and\
  \citenamefont {Fangohr}}]{Bisotti2018}%
  \BibitemOpen
  \bibfield  {author} {\bibinfo {author} {\bibfnamefont {M.-A.}\ \bibnamefont
  {Bisotti}}, \bibinfo {author} {\bibfnamefont {D.}~\bibnamefont
  {{Cort{\'e}s-Ortu{\~n}o}}}, \bibinfo {author} {\bibfnamefont
  {R.}~\bibnamefont {Pepper}}, \bibinfo {author} {\bibfnamefont
  {W.}~\bibnamefont {Wang}}, \bibinfo {author} {\bibfnamefont {M.}~\bibnamefont
  {Beg}}, \bibinfo {author} {\bibfnamefont {T.}~\bibnamefont {Kluyver}},\ and\
  \bibinfo {author} {\bibfnamefont {H.}~\bibnamefont {Fangohr}},\ }\bibfield
  {journal} {\bibinfo  {journal} {J. Open Res. Softw.}\ }\textbf {\bibinfo
  {volume} {6}},\ \href {https://doi.org/10.5334/jors.223} {10.5334/jors.223}
  (\bibinfo {year} {2018})\BibitemShut {NoStop}%
\bibitem [{\citenamefont {Scholz}\ \emph {et~al.}(2003)\citenamefont {Scholz},
  \citenamefont {Fidler}, \citenamefont {Schrefl}, \citenamefont {Suess},
  \citenamefont {Dittrich}, \citenamefont {Forster},\ and\ \citenamefont
  {Tsiantos}}]{Scholz2003}%
  \BibitemOpen
  \bibfield  {author} {\bibinfo {author} {\bibfnamefont {W.}~\bibnamefont
  {Scholz}}, \bibinfo {author} {\bibfnamefont {J.}~\bibnamefont {Fidler}},
  \bibinfo {author} {\bibfnamefont {T.}~\bibnamefont {Schrefl}}, \bibinfo
  {author} {\bibfnamefont {D.}~\bibnamefont {Suess}}, \bibinfo {author}
  {\bibfnamefont {R.}~\bibnamefont {Dittrich}}, \bibinfo {author}
  {\bibfnamefont {H.}~\bibnamefont {Forster}},\ and\ \bibinfo {author}
  {\bibfnamefont {V.}~\bibnamefont {Tsiantos}},\ }\href
  {https://doi.org/10.1016/S0927-0256(03)00119-8} {\bibfield  {journal}
  {\bibinfo  {journal} {Computational Materials Science}\ }\textbf {\bibinfo
  {volume} {28}},\ \bibinfo {pages} {366} (\bibinfo {year} {2003})}\BibitemShut
  {NoStop}%
\bibitem [{\citenamefont {Fischbacher}\ \emph {et~al.}(2007)\citenamefont
  {Fischbacher}, \citenamefont {Franchin}, \citenamefont {Bordignon},\ and\
  \citenamefont {Fangohr}}]{Fischbacher2007}%
  \BibitemOpen
  \bibfield  {author} {\bibinfo {author} {\bibfnamefont {T.}~\bibnamefont
  {Fischbacher}}, \bibinfo {author} {\bibfnamefont {M.}~\bibnamefont
  {Franchin}}, \bibinfo {author} {\bibfnamefont {G.}~\bibnamefont
  {Bordignon}},\ and\ \bibinfo {author} {\bibfnamefont {H.}~\bibnamefont
  {Fangohr}},\ }\href {https://doi.org/10.1109/TMAG.2007.893843} {\bibfield
  {journal} {\bibinfo  {journal} {IEEE Transactions on Magnetics}\ }\textbf
  {\bibinfo {volume} {43}},\ \bibinfo {pages} {2896} (\bibinfo {year}
  {2007})}\BibitemShut {NoStop}%
\bibitem [{\citenamefont {Pfeiler}\ \emph {et~al.}(2020)\citenamefont
  {Pfeiler}, \citenamefont {Ruggeri}, \citenamefont {Stiftner}, \citenamefont
  {Exl}, \citenamefont {Hochsteger}, \citenamefont {Hrkac}, \citenamefont
  {Sch{\"o}berl}, \citenamefont {Mauser},\ and\ \citenamefont
  {Praetorius}}]{Pfeiler2020}%
  \BibitemOpen
  \bibfield  {author} {\bibinfo {author} {\bibfnamefont {C.-M.}\ \bibnamefont
  {Pfeiler}}, \bibinfo {author} {\bibfnamefont {M.}~\bibnamefont {Ruggeri}},
  \bibinfo {author} {\bibfnamefont {B.}~\bibnamefont {Stiftner}}, \bibinfo
  {author} {\bibfnamefont {L.}~\bibnamefont {Exl}}, \bibinfo {author}
  {\bibfnamefont {M.}~\bibnamefont {Hochsteger}}, \bibinfo {author}
  {\bibfnamefont {G.}~\bibnamefont {Hrkac}}, \bibinfo {author} {\bibfnamefont
  {J.}~\bibnamefont {Sch{\"o}berl}}, \bibinfo {author} {\bibfnamefont {N.~J.}\
  \bibnamefont {Mauser}},\ and\ \bibinfo {author} {\bibfnamefont
  {D.}~\bibnamefont {Praetorius}},\ }\href
  {https://doi.org/10.1016/j.cpc.2019.106965} {\bibfield  {journal} {\bibinfo
  {journal} {Computer Physics Communications}\ }\textbf {\bibinfo {volume}
  {248}},\ \bibinfo {pages} {106965} (\bibinfo {year} {2020})}\BibitemShut
  {NoStop}%
\bibitem [{\citenamefont {Bruckner}\ \emph {et~al.}(2023)\citenamefont
  {Bruckner}, \citenamefont {Koraltan}, \citenamefont {Abert},\ and\
  \citenamefont {Suess}}]{Bruckner2023}%
  \BibitemOpen
  \bibfield  {author} {\bibinfo {author} {\bibfnamefont {F.}~\bibnamefont
  {Bruckner}}, \bibinfo {author} {\bibfnamefont {S.}~\bibnamefont {Koraltan}},
  \bibinfo {author} {\bibfnamefont {C.}~\bibnamefont {Abert}},\ and\ \bibinfo
  {author} {\bibfnamefont {D.}~\bibnamefont {Suess}},\ }\href
  {https://doi.org/10.1038/s41598-023-39192-5} {\bibfield  {journal} {\bibinfo
  {journal} {Scientific Reports}\ }\textbf {\bibinfo {volume} {13}},\ \bibinfo
  {pages} {12054} (\bibinfo {year} {2023})}\BibitemShut {NoStop}%
\bibitem [{\citenamefont {Besard}\ \emph {et~al.}(2019)\citenamefont {Besard},
  \citenamefont {Churavy}, \citenamefont {Edelman},\ and\ \citenamefont
  {Sutter}}]{Besard2019}%
  \BibitemOpen
  \bibfield  {author} {\bibinfo {author} {\bibfnamefont {T.}~\bibnamefont
  {Besard}}, \bibinfo {author} {\bibfnamefont {V.}~\bibnamefont {Churavy}},
  \bibinfo {author} {\bibfnamefont {A.}~\bibnamefont {Edelman}},\ and\ \bibinfo
  {author} {\bibfnamefont {B.~D.}\ \bibnamefont {Sutter}},\ }\href
  {https://doi.org/10.1016/j.advengsoft.2019.02.002} {\bibfield  {journal}
  {\bibinfo  {journal} {Advances in Engineering Software}\ }\textbf {\bibinfo
  {volume} {132}},\ \bibinfo {pages} {29} (\bibinfo {year} {2019})}\BibitemShut
  {NoStop}%
\bibitem [{\citenamefont {Churavy}\ \emph {et~al.}(2024)\citenamefont
  {Churavy}, \citenamefont {Aluthge}, \citenamefont {Smirnov}, \citenamefont
  {Schloss}, \citenamefont {Samaroo}, \citenamefont {Wilcox}, \citenamefont
  {Byrne}, \citenamefont {Besard}, \citenamefont {Ramadhan}, \citenamefont
  {Waruszewski}, \citenamefont {Schaub}, \citenamefont {Meredith},
  \citenamefont {Moses}, \citenamefont {Bolewski}, \citenamefont
  {Constantinou}, \citenamefont {Ng}, \citenamefont {Bauer}, \citenamefont
  {Schanen}, \citenamefont {{johnbcoughlin}}, \citenamefont {Shah},
  \citenamefont {Dixit}, \citenamefont {Chor}, \citenamefont {Holy},
  \citenamefont {Arakaki}, \citenamefont {Yalburgi}, \citenamefont {Liu},\ and\
  \citenamefont {Haraldsson}}]{ValentinChuravy2024}%
  \BibitemOpen
  \bibfield  {author} {\bibinfo {author} {\bibfnamefont {V.}~\bibnamefont
  {Churavy}}, \bibinfo {author} {\bibfnamefont {D.}~\bibnamefont {Aluthge}},
  \bibinfo {author} {\bibfnamefont {A.}~\bibnamefont {Smirnov}}, \bibinfo
  {author} {\bibfnamefont {J.}~\bibnamefont {Schloss}}, \bibinfo {author}
  {\bibfnamefont {J.}~\bibnamefont {Samaroo}}, \bibinfo {author} {\bibfnamefont
  {L.~C.}\ \bibnamefont {Wilcox}}, \bibinfo {author} {\bibfnamefont
  {S.}~\bibnamefont {Byrne}}, \bibinfo {author} {\bibfnamefont
  {T.}~\bibnamefont {Besard}}, \bibinfo {author} {\bibfnamefont
  {A.}~\bibnamefont {Ramadhan}}, \bibinfo {author} {\bibfnamefont
  {M.}~\bibnamefont {Waruszewski}}, \bibinfo {author} {\bibfnamefont {S.~D.}\
  \bibnamefont {Schaub}}, \bibinfo {author} {\bibnamefont {Meredith}}, \bibinfo
  {author} {\bibfnamefont {W.}~\bibnamefont {Moses}}, \bibinfo {author}
  {\bibfnamefont {J.}~\bibnamefont {Bolewski}}, \bibinfo {author}
  {\bibfnamefont {N.~C.}\ \bibnamefont {Constantinou}}, \bibinfo {author}
  {\bibfnamefont {M.}~\bibnamefont {Ng}}, \bibinfo {author} {\bibfnamefont
  {C.}~\bibnamefont {Bauer}}, \bibinfo {author} {\bibfnamefont
  {M.}~\bibnamefont {Schanen}}, \bibinfo {author} {\bibnamefont
  {{johnbcoughlin}}}, \bibinfo {author} {\bibfnamefont {V.~B.}\ \bibnamefont
  {Shah}}, \bibinfo {author} {\bibfnamefont {V.~K.}\ \bibnamefont {Dixit}},
  \bibinfo {author} {\bibfnamefont {T.}~\bibnamefont {Chor}}, \bibinfo {author}
  {\bibfnamefont {T.}~\bibnamefont {Holy}}, \bibinfo {author} {\bibfnamefont
  {T.}~\bibnamefont {Arakaki}}, \bibinfo {author} {\bibfnamefont
  {S.}~\bibnamefont {Yalburgi}}, \bibinfo {author} {\bibfnamefont
  {R.}~\bibnamefont {Liu}},\ and\ \bibinfo {author} {\bibfnamefont
  {P.}~\bibnamefont {Haraldsson}},\ }\href
  {https://doi.org/10.5281/ZENODO.10780635} {\bibinfo {title}
  {{{JuliaGPU}}/{{KernelAbstractions}}.jl: V0.9.18}},\ \bibinfo {howpublished}
  {[object Object]} (\bibinfo {year} {2024})\BibitemShut {NoStop}%
\bibitem [{\citenamefont {{Marc-Antonio Bisotti}}\ \emph
  {et~al.}(2018)\citenamefont {{Marc-Antonio Bisotti}}, \citenamefont {Beg},
  \citenamefont {{Weiwei Wang}}, \citenamefont {Albert}, \citenamefont
  {Chernyshenko}, \citenamefont {{Cort{\'e}s-Ortu{\~n}o}}, \citenamefont
  {Pepper}, \citenamefont {Vousden}, \citenamefont {Carey}, \citenamefont
  {Fuchs}, \citenamefont {Johansen}, \citenamefont {Balaban}, \citenamefont
  {Breth}, \citenamefont {Kluyver},\ and\ \citenamefont
  {Fangohr}}]{Marc-AntonioBisotti2018}%
  \BibitemOpen
  \bibfield  {author} {\bibinfo {author} {\bibnamefont {{Marc-Antonio
  Bisotti}}}, \bibinfo {author} {\bibfnamefont {M.}~\bibnamefont {Beg}},
  \bibinfo {author} {\bibnamefont {{Weiwei Wang}}}, \bibinfo {author}
  {\bibfnamefont {M.}~\bibnamefont {Albert}}, \bibinfo {author} {\bibfnamefont
  {D.}~\bibnamefont {Chernyshenko}}, \bibinfo {author} {\bibfnamefont
  {D.}~\bibnamefont {{Cort{\'e}s-Ortu{\~n}o}}}, \bibinfo {author}
  {\bibfnamefont {R.~A.}\ \bibnamefont {Pepper}}, \bibinfo {author}
  {\bibfnamefont {M.}~\bibnamefont {Vousden}}, \bibinfo {author} {\bibfnamefont
  {R.}~\bibnamefont {Carey}}, \bibinfo {author} {\bibfnamefont
  {H.}~\bibnamefont {Fuchs}}, \bibinfo {author} {\bibfnamefont
  {A.}~\bibnamefont {Johansen}}, \bibinfo {author} {\bibfnamefont
  {G.}~\bibnamefont {Balaban}}, \bibinfo {author} {\bibfnamefont
  {L.}~\bibnamefont {Breth}}, \bibinfo {author} {\bibfnamefont
  {T.}~\bibnamefont {Kluyver}},\ and\ \bibinfo {author} {\bibfnamefont
  {H.}~\bibnamefont {Fangohr}},\ }\href
  {https://doi.org/10.5281/ZENODO.1216011} {\bibinfo {title} {{{FinMag}}:
  Finite-element micromagnetic simulation tool}},\ \bibinfo {howpublished}
  {Zenodo} (\bibinfo {year} {2018})\BibitemShut {NoStop}%
\bibitem [{\citenamefont {Newell}\ \emph {et~al.}(1993)\citenamefont {Newell},
  \citenamefont {Williams},\ and\ \citenamefont {Dunlop}}]{Newell1993}%
  \BibitemOpen
  \bibfield  {author} {\bibinfo {author} {\bibfnamefont {A.~J.}\ \bibnamefont
  {Newell}}, \bibinfo {author} {\bibfnamefont {W.}~\bibnamefont {Williams}},\
  and\ \bibinfo {author} {\bibfnamefont {D.~J.}\ \bibnamefont {Dunlop}},\
  }\href {https://doi.org/10.1029/93JB00694} {\bibfield  {journal} {\bibinfo
  {journal} {Journal of Geophysical Research}\ }\textbf {\bibinfo {volume}
  {98}},\ \bibinfo {pages} {9551} (\bibinfo {year} {1993})}\BibitemShut
  {NoStop}%
\bibitem [{\citenamefont {Abert}\ \emph {et~al.}(2015)\citenamefont {Abert},
  \citenamefont {Bruckner}, \citenamefont {Vogler}, \citenamefont {Windl},
  \citenamefont {Thanhoffer},\ and\ \citenamefont {Suess}}]{Abert2015}%
  \BibitemOpen
  \bibfield  {author} {\bibinfo {author} {\bibfnamefont {C.}~\bibnamefont
  {Abert}}, \bibinfo {author} {\bibfnamefont {F.}~\bibnamefont {Bruckner}},
  \bibinfo {author} {\bibfnamefont {C.}~\bibnamefont {Vogler}}, \bibinfo
  {author} {\bibfnamefont {R.}~\bibnamefont {Windl}}, \bibinfo {author}
  {\bibfnamefont {R.}~\bibnamefont {Thanhoffer}},\ and\ \bibinfo {author}
  {\bibfnamefont {D.}~\bibnamefont {Suess}},\ }\href
  {https://doi.org/10.1016/j.jmmm.2015.03.081} {\bibfield  {journal} {\bibinfo
  {journal} {Journal of Magnetism and Magnetic Materials}\ }\textbf {\bibinfo
  {volume} {387}},\ \bibinfo {pages} {13} (\bibinfo {year} {2015})},\ \Eprint
  {https://arxiv.org/abs/1411.7188} {arxiv:1411.7188 [physics]} \BibitemShut
  {NoStop}%
\bibitem [{\citenamefont {Wang}(2015)}]{Wang2015}%
  \BibitemOpen
  \bibfield  {author} {\bibinfo {author} {\bibfnamefont {W.}~\bibnamefont
  {Wang}},\ }\emph {\bibinfo {title} {Computer Simulation Studies of Complex
  Magnetic Materials}},\ \href@noop {} {Ph.D. thesis},\ \bibinfo  {school}
  {University of Southampton} (\bibinfo {year} {2015})\BibitemShut {NoStop}%
\bibitem [{\citenamefont {Abert}\ \emph {et~al.}(2012)\citenamefont {Abert},
  \citenamefont {Selke}, \citenamefont {Kruger},\ and\ \citenamefont
  {Drews}}]{Abert2012}%
  \BibitemOpen
  \bibfield  {author} {\bibinfo {author} {\bibfnamefont {C.}~\bibnamefont
  {Abert}}, \bibinfo {author} {\bibfnamefont {G.}~\bibnamefont {Selke}},
  \bibinfo {author} {\bibfnamefont {B.}~\bibnamefont {Kruger}},\ and\ \bibinfo
  {author} {\bibfnamefont {A.}~\bibnamefont {Drews}},\ }\href
  {https://doi.org/10.1109/TMAG.2011.2172806} {\bibfield  {journal} {\bibinfo
  {journal} {IEEE Transactions on Magnetics}\ }\textbf {\bibinfo {volume}
  {48}},\ \bibinfo {pages} {1105} (\bibinfo {year} {2012})}\BibitemShut
  {NoStop}%
\bibitem [{\citenamefont {M{\"u}hlbauer}\ \emph {et~al.}(2009)\citenamefont
  {M{\"u}hlbauer}, \citenamefont {Binz}, \citenamefont {Jonietz}, \citenamefont
  {Pfleiderer}, \citenamefont {Rosch}, \citenamefont {Neubauer}, \citenamefont
  {Georgii},\ and\ \citenamefont {B{\"o}ni}}]{Muhlbauer2009}%
  \BibitemOpen
  \bibfield  {author} {\bibinfo {author} {\bibfnamefont {S.}~\bibnamefont
  {M{\"u}hlbauer}}, \bibinfo {author} {\bibfnamefont {B.}~\bibnamefont {Binz}},
  \bibinfo {author} {\bibfnamefont {F.}~\bibnamefont {Jonietz}}, \bibinfo
  {author} {\bibfnamefont {C.}~\bibnamefont {Pfleiderer}}, \bibinfo {author}
  {\bibfnamefont {A.}~\bibnamefont {Rosch}}, \bibinfo {author} {\bibfnamefont
  {A.}~\bibnamefont {Neubauer}}, \bibinfo {author} {\bibfnamefont
  {R.}~\bibnamefont {Georgii}},\ and\ \bibinfo {author} {\bibfnamefont
  {P.}~\bibnamefont {B{\"o}ni}},\ }\href
  {https://doi.org/10.1126/science.1166767} {\bibfield  {journal} {\bibinfo
  {journal} {Science}\ }\textbf {\bibinfo {volume} {323}},\ \bibinfo {pages}
  {915} (\bibinfo {year} {2009})}\BibitemShut {NoStop}%
\bibitem [{\citenamefont {Huang}\ and\ \citenamefont
  {Chien}(2012)}]{Huang2012}%
  \BibitemOpen
  \bibfield  {author} {\bibinfo {author} {\bibfnamefont {S.~X.}\ \bibnamefont
  {Huang}}\ and\ \bibinfo {author} {\bibfnamefont {C.~L.}\ \bibnamefont
  {Chien}},\ }\href {https://doi.org/10.1103/PhysRevLett.108.267201} {\bibfield
   {journal} {\bibinfo  {journal} {Physical Review Letters}\ }\textbf {\bibinfo
  {volume} {108}},\ \bibinfo {pages} {267201} (\bibinfo {year}
  {2012})}\BibitemShut {NoStop}%
\bibitem [{\citenamefont {{Cort{\'e}s-Ortu{\~n}o}}\ \emph
  {et~al.}(2018)\citenamefont {{Cort{\'e}s-Ortu{\~n}o}}, \citenamefont {Beg},
  \citenamefont {Nehruji}, \citenamefont {Breth}, \citenamefont {Pepper},
  \citenamefont {Kluyver}, \citenamefont {Downing}, \citenamefont {Hesjedal},
  \citenamefont {Hatton}, \citenamefont {Lancaster}, \citenamefont {Hertel},
  \citenamefont {Hovorka},\ and\ \citenamefont {Fangohr}}]{Cortes-Ortuno2018}%
  \BibitemOpen
  \bibfield  {author} {\bibinfo {author} {\bibfnamefont {D.}~\bibnamefont
  {{Cort{\'e}s-Ortu{\~n}o}}}, \bibinfo {author} {\bibfnamefont
  {M.}~\bibnamefont {Beg}}, \bibinfo {author} {\bibfnamefont {V.}~\bibnamefont
  {Nehruji}}, \bibinfo {author} {\bibfnamefont {L.}~\bibnamefont {Breth}},
  \bibinfo {author} {\bibfnamefont {R.}~\bibnamefont {Pepper}}, \bibinfo
  {author} {\bibfnamefont {T.}~\bibnamefont {Kluyver}}, \bibinfo {author}
  {\bibfnamefont {G.}~\bibnamefont {Downing}}, \bibinfo {author} {\bibfnamefont
  {T.}~\bibnamefont {Hesjedal}}, \bibinfo {author} {\bibfnamefont
  {P.}~\bibnamefont {Hatton}}, \bibinfo {author} {\bibfnamefont
  {T.}~\bibnamefont {Lancaster}}, \bibinfo {author} {\bibfnamefont
  {R.}~\bibnamefont {Hertel}}, \bibinfo {author} {\bibfnamefont
  {O.}~\bibnamefont {Hovorka}},\ and\ \bibinfo {author} {\bibfnamefont
  {H.}~\bibnamefont {Fangohr}},\ }\href
  {https://doi.org/10.1088/1367-2630/aaea1c} {\bibfield  {journal} {\bibinfo
  {journal} {New Journal of Physics}\ }\textbf {\bibinfo {volume} {20}},\
  \bibinfo {pages} {113015} (\bibinfo {year} {2018})}\BibitemShut {NoStop}%
\bibitem [{\citenamefont {Vedmedenko}\ \emph {et~al.}(2019)\citenamefont
  {Vedmedenko}, \citenamefont {Riego}, \citenamefont {Arregi},\ and\
  \citenamefont {Berger}}]{Vedmedenko2019}%
  \BibitemOpen
  \bibfield  {author} {\bibinfo {author} {\bibfnamefont {E.~Y.}\ \bibnamefont
  {Vedmedenko}}, \bibinfo {author} {\bibfnamefont {P.}~\bibnamefont {Riego}},
  \bibinfo {author} {\bibfnamefont {J.~A.}\ \bibnamefont {Arregi}},\ and\
  \bibinfo {author} {\bibfnamefont {A.}~\bibnamefont {Berger}},\ }\href
  {https://doi.org/10.1103/PhysRevLett.122.257202} {\bibfield  {journal}
  {\bibinfo  {journal} {Physical Review Letters}\ }\textbf {\bibinfo {volume}
  {122}},\ \bibinfo {pages} {257202} (\bibinfo {year} {2019})}\BibitemShut
  {NoStop}%
\bibitem [{\citenamefont {Han}\ \emph {et~al.}(2019)\citenamefont {Han},
  \citenamefont {Lee}, \citenamefont {Hanke}, \citenamefont {Mokrousov},
  \citenamefont {Kim}, \citenamefont {Yoo}, \citenamefont {Van~Hees},
  \citenamefont {Kim}, \citenamefont {Lavrijsen}, \citenamefont {You},
  \citenamefont {Swagten}, \citenamefont {Jung},\ and\ \citenamefont
  {Kl{\"a}ui}}]{Han2019}%
  \BibitemOpen
  \bibfield  {author} {\bibinfo {author} {\bibfnamefont {D.-S.}\ \bibnamefont
  {Han}}, \bibinfo {author} {\bibfnamefont {K.}~\bibnamefont {Lee}}, \bibinfo
  {author} {\bibfnamefont {J.-P.}\ \bibnamefont {Hanke}}, \bibinfo {author}
  {\bibfnamefont {Y.}~\bibnamefont {Mokrousov}}, \bibinfo {author}
  {\bibfnamefont {K.-W.}\ \bibnamefont {Kim}}, \bibinfo {author} {\bibfnamefont
  {W.}~\bibnamefont {Yoo}}, \bibinfo {author} {\bibfnamefont {Y.~L.~W.}\
  \bibnamefont {Van~Hees}}, \bibinfo {author} {\bibfnamefont {T.-W.}\
  \bibnamefont {Kim}}, \bibinfo {author} {\bibfnamefont {R.}~\bibnamefont
  {Lavrijsen}}, \bibinfo {author} {\bibfnamefont {C.-Y.}\ \bibnamefont {You}},
  \bibinfo {author} {\bibfnamefont {H.~J.~M.}\ \bibnamefont {Swagten}},
  \bibinfo {author} {\bibfnamefont {M.-H.}\ \bibnamefont {Jung}},\ and\
  \bibinfo {author} {\bibfnamefont {M.}~\bibnamefont {Kl{\"a}ui}},\ }\href
  {https://doi.org/10.1038/s41563-019-0370-z} {\bibfield  {journal} {\bibinfo
  {journal} {Nature Materials}\ }\textbf {\bibinfo {volume} {18}},\ \bibinfo
  {pages} {703} (\bibinfo {year} {2019})}\BibitemShut {NoStop}%
\bibitem [{\citenamefont {Tatara}\ \emph {et~al.}(2008)\citenamefont {Tatara},
  \citenamefont {Kohno},\ and\ \citenamefont {Shibata}}]{Tatara2008}%
  \BibitemOpen
  \bibfield  {author} {\bibinfo {author} {\bibfnamefont {G.}~\bibnamefont
  {Tatara}}, \bibinfo {author} {\bibfnamefont {H.}~\bibnamefont {Kohno}},\ and\
  \bibinfo {author} {\bibfnamefont {J.}~\bibnamefont {Shibata}},\ }\href
  {https://doi.org/10.1016/j.physrep.2008.07.003} {\bibfield  {journal}
  {\bibinfo  {journal} {Physics Reports}\ }\textbf {\bibinfo {volume} {468}},\
  \bibinfo {pages} {213} (\bibinfo {year} {2008})},\ \Eprint
  {https://arxiv.org/abs/0807.2894} {arxiv:0807.2894} \BibitemShut {NoStop}%
\bibitem [{\citenamefont {Nowak}(2007)}]{Nowak2007}%
  \BibitemOpen
  \bibfield  {author} {\bibinfo {author} {\bibfnamefont {U.}~\bibnamefont
  {Nowak}},\ }in\ \href {https://doi.org/10.1002/9780470022184.hmm205} {\emph
  {\bibinfo {booktitle} {Handbook of {{Magnetism}} and {{Advanced Magnetic
  Materials}}}}},\ \bibinfo {editor} {edited by\ \bibinfo {editor}
  {\bibfnamefont {H.}~\bibnamefont {Kronm{\"u}ller}}\ and\ \bibinfo {editor}
  {\bibfnamefont {S.}~\bibnamefont {Parkin}}}\ (\bibinfo  {publisher} {Wiley},\
  \bibinfo {year} {2007})\ \bibinfo {edition} {1st}\ ed.\BibitemShut {Stop}%
\bibitem [{\citenamefont {Skubic}\ \emph {et~al.}(2008)\citenamefont {Skubic},
  \citenamefont {Hellsvik}, \citenamefont {Nordstr{\"o}m},\ and\ \citenamefont
  {Eriksson}}]{Skubic2008}%
  \BibitemOpen
  \bibfield  {author} {\bibinfo {author} {\bibfnamefont {B.}~\bibnamefont
  {Skubic}}, \bibinfo {author} {\bibfnamefont {J.}~\bibnamefont {Hellsvik}},
  \bibinfo {author} {\bibfnamefont {L.}~\bibnamefont {Nordstr{\"o}m}},\ and\
  \bibinfo {author} {\bibfnamefont {O.}~\bibnamefont {Eriksson}},\ }\href
  {https://doi.org/10.1088/0953-8984/20/31/315203} {\bibfield  {journal}
  {\bibinfo  {journal} {Journal of Physics: Condensed Matter}\ }\textbf
  {\bibinfo {volume} {20}},\ \bibinfo {pages} {14} (\bibinfo {year} {2008})},\
  \Eprint {https://arxiv.org/abs/0806.1582} {arxiv:0806.1582} \BibitemShut
  {NoStop}%
\bibitem [{\citenamefont {Evans}\ \emph {et~al.}(2014)\citenamefont {Evans},
  \citenamefont {Fan}, \citenamefont {Chureemart}, \citenamefont {a~Ostler},
  \citenamefont {a~Ellis},\ and\ \citenamefont {Chantrell}}]{Evans2014}%
  \BibitemOpen
  \bibfield  {author} {\bibinfo {author} {\bibfnamefont {R.~F.~L.}\
  \bibnamefont {Evans}}, \bibinfo {author} {\bibfnamefont {W.~J.}\ \bibnamefont
  {Fan}}, \bibinfo {author} {\bibfnamefont {P.}~\bibnamefont {Chureemart}},
  \bibinfo {author} {\bibfnamefont {T.}~\bibnamefont {a~Ostler}}, \bibinfo
  {author} {\bibfnamefont {M.~O.}\ \bibnamefont {a~Ellis}},\ and\ \bibinfo
  {author} {\bibfnamefont {R.~W.}\ \bibnamefont {Chantrell}},\ }\href
  {https://doi.org/10.1088/0953-8984/26/10/103202} {\bibfield  {journal}
  {\bibinfo  {journal} {Journal of physics. Condensed matter}\ }\textbf
  {\bibinfo {volume} {26}},\ \bibinfo {pages} {103202} (\bibinfo {year}
  {2014})},\ \Eprint {https://arxiv.org/abs/24552692} {24552692} \BibitemShut
  {NoStop}%
\bibitem [{\citenamefont {Dzyaloshinskii}(1958)}]{Dzyaloshinskii1958}%
  \BibitemOpen
  \bibfield  {author} {\bibinfo {author} {\bibfnamefont {I.}~\bibnamefont
  {Dzyaloshinskii}},\ }\href@noop {} {\bibfield  {journal} {\bibinfo  {journal}
  {J. Phys. Chem. Solids}\ }\textbf {\bibinfo {volume} {4}},\ \bibinfo {pages}
  {241} (\bibinfo {year} {1958})}\BibitemShut {NoStop}%
\bibitem [{\citenamefont {Moriya}(1960)}]{Moriya1960}%
  \BibitemOpen
  \bibfield  {author} {\bibinfo {author} {\bibfnamefont {T.}~\bibnamefont
  {Moriya}},\ }\href@noop {} {\bibfield  {journal} {\bibinfo  {journal} {Phys.
  Rev.}\ }\textbf {\bibinfo {volume} {120}},\ \bibinfo {pages} {91} (\bibinfo
  {year} {1960})}\BibitemShut {NoStop}%
\bibitem [{\citenamefont {Rohart}\ and\ \citenamefont
  {Thiaville}(2013)}]{Rohart2013}%
  \BibitemOpen
  \bibfield  {author} {\bibinfo {author} {\bibfnamefont {S.}~\bibnamefont
  {Rohart}}\ and\ \bibinfo {author} {\bibfnamefont {A.}~\bibnamefont
  {Thiaville}},\ }\href {https://doi.org/10.1103/PhysRevB.88.184422} {\bibfield
   {journal} {\bibinfo  {journal} {Physical Review B}\ }\textbf {\bibinfo
  {volume} {88}},\ \bibinfo {pages} {184422} (\bibinfo {year} {2013})},\
  \Eprint {https://arxiv.org/abs/1310.0666} {arxiv:1310.0666} \BibitemShut
  {NoStop}%
\bibitem [{\citenamefont {Gilbert}(2004)}]{Gilbert2004}%
  \BibitemOpen
  \bibfield  {author} {\bibinfo {author} {\bibfnamefont {T.~L.}\ \bibnamefont
  {Gilbert}},\ }\href {https://doi.org/10.1109/TMAG.2004.836740} {\bibfield
  {journal} {\bibinfo  {journal} {IEEE Transactions on Magnetics}\ }\textbf
  {\bibinfo {volume} {40}},\ \bibinfo {pages} {3443} (\bibinfo {year}
  {2004})}\BibitemShut {NoStop}%
\bibitem [{\citenamefont {Press}\ and\ \citenamefont
  {Teukolsky}(1992)}]{Press1992}%
  \BibitemOpen
  \bibfield  {author} {\bibinfo {author} {\bibfnamefont {W.~H.}\ \bibnamefont
  {Press}}\ and\ \bibinfo {author} {\bibfnamefont {S.~A.}\ \bibnamefont
  {Teukolsky}},\ }\href {https://doi.org/10.1063/1.4823060} {\bibfield
  {journal} {\bibinfo  {journal} {Computers in Physics}\ }\textbf {\bibinfo
  {volume} {6}},\ \bibinfo {pages} {188} (\bibinfo {year} {1992})}\BibitemShut
  {NoStop}%
\bibitem [{\citenamefont {Slonczewski}(1996)}]{Slonczewski1996}%
  \BibitemOpen
  \bibfield  {author} {\bibinfo {author} {\bibfnamefont {J.}~\bibnamefont
  {Slonczewski}},\ }\href {https://doi.org/10.1016/0304-8853(96)00062-5}
  {\bibfield  {journal} {\bibinfo  {journal} {Journal of Magnetism and Magnetic
  Materials}\ }\textbf {\bibinfo {volume} {159}},\ \bibinfo {pages} {L1}
  (\bibinfo {year} {1996})}\BibitemShut {NoStop}%
\bibitem [{\citenamefont {Zhang}\ and\ \citenamefont {Li}(2004)}]{Zhang2004}%
  \BibitemOpen
  \bibfield  {author} {\bibinfo {author} {\bibfnamefont {S.}~\bibnamefont
  {Zhang}}\ and\ \bibinfo {author} {\bibfnamefont {Z.}~\bibnamefont {Li}},\
  }\href {https://doi.org/10.1103/PhysRevLett.93.127204} {\bibfield  {journal}
  {\bibinfo  {journal} {Physical Review Letters}\ }\textbf {\bibinfo {volume}
  {93}},\ \bibinfo {pages} {127204} (\bibinfo {year} {2004})}\BibitemShut
  {NoStop}%
\bibitem [{\citenamefont {Meo}\ \emph {et~al.}(2023)\citenamefont {Meo},
  \citenamefont {Cronshaw}, \citenamefont {Jenkins}, \citenamefont {Lees},\
  and\ \citenamefont {Evans}}]{Meo2023}%
  \BibitemOpen
  \bibfield  {author} {\bibinfo {author} {\bibfnamefont {A.}~\bibnamefont
  {Meo}}, \bibinfo {author} {\bibfnamefont {C.~E.}\ \bibnamefont {Cronshaw}},
  \bibinfo {author} {\bibfnamefont {S.}~\bibnamefont {Jenkins}}, \bibinfo
  {author} {\bibfnamefont {A.}~\bibnamefont {Lees}},\ and\ \bibinfo {author}
  {\bibfnamefont {R.~F.~L.}\ \bibnamefont {Evans}},\ }\href
  {https://doi.org/10.1088/1361-648X/ac9c80} {\bibfield  {journal} {\bibinfo
  {journal} {Journal of Physics: Condensed Matter}\ }\textbf {\bibinfo {volume}
  {35}},\ \bibinfo {pages} {025801} (\bibinfo {year} {2023})}\BibitemShut
  {NoStop}%
\bibitem [{\citenamefont {Krishnaprasad}\ and\ \citenamefont
  {Tan}(2001)}]{Krishnaprasad2001a}%
  \BibitemOpen
  \bibfield  {author} {\bibinfo {author} {\bibfnamefont {P.~S.}\ \bibnamefont
  {Krishnaprasad}}\ and\ \bibinfo {author} {\bibfnamefont {X.}~\bibnamefont
  {Tan}},\ }\href {https://doi.org/10.1016/S0921-4526(01)01003-1} {\bibfield
  {journal} {\bibinfo  {journal} {Physica B: Condensed Matter}\ }\textbf
  {\bibinfo {volume} {306}},\ \bibinfo {pages} {195} (\bibinfo {year}
  {2001})}\BibitemShut {NoStop}%
\bibitem [{\citenamefont {Iserles}\ and\ \citenamefont
  {Zanna}(2000)}]{Iserles2000}%
  \BibitemOpen
  \bibfield  {author} {\bibinfo {author} {\bibfnamefont {A.}~\bibnamefont
  {Iserles}}\ and\ \bibinfo {author} {\bibfnamefont {A.}~\bibnamefont
  {Zanna}},\ }\href {https://doi.org/10.1112/S1461157000000206} {\bibfield
  {journal} {\bibinfo  {journal} {LMS Journal of Computation and Mathematics}\
  }\textbf {\bibinfo {volume} {3}},\ \bibinfo {pages} {44} (\bibinfo {year}
  {2000})}\BibitemShut {NoStop}%
\bibitem [{\citenamefont {Diele}\ \emph {et~al.}(1998)\citenamefont {Diele},
  \citenamefont {Lopez},\ and\ \citenamefont {Peluso}}]{Diele1998a}%
  \BibitemOpen
  \bibfield  {author} {\bibinfo {author} {\bibfnamefont {F.}~\bibnamefont
  {Diele}}, \bibinfo {author} {\bibfnamefont {L.}~\bibnamefont {Lopez}},\ and\
  \bibinfo {author} {\bibfnamefont {R.}~\bibnamefont {Peluso}},\ }\href
  {https://doi.org/10.1023/A:1018908700358} {\bibfield  {journal} {\bibinfo
  {journal} {Advances in Computational Mathematics}\ }\textbf {\bibinfo
  {volume} {8}},\ \bibinfo {pages} {317} (\bibinfo {year} {1998})}\BibitemShut
  {NoStop}%
\bibitem [{\citenamefont {Abert}\ \emph {et~al.}(2014)\citenamefont {Abert},
  \citenamefont {Wautischer}, \citenamefont {Bruckner}, \citenamefont {Satz},\
  and\ \citenamefont {Suess}}]{Abert2014a}%
  \BibitemOpen
  \bibfield  {author} {\bibinfo {author} {\bibfnamefont {C.}~\bibnamefont
  {Abert}}, \bibinfo {author} {\bibfnamefont {G.}~\bibnamefont {Wautischer}},
  \bibinfo {author} {\bibfnamefont {F.}~\bibnamefont {Bruckner}}, \bibinfo
  {author} {\bibfnamefont {A.}~\bibnamefont {Satz}},\ and\ \bibinfo {author}
  {\bibfnamefont {D.}~\bibnamefont {Suess}},\ }\href
  {https://doi.org/10.1063/1.4896360} {\bibfield  {journal} {\bibinfo
  {journal} {Journal of Applied Physics}\ }\textbf {\bibinfo {volume} {116}},\
  \bibinfo {pages} {123908} (\bibinfo {year} {2014})}\BibitemShut {NoStop}%
\bibitem [{\citenamefont {Exl}\ \emph {et~al.}(2014)\citenamefont {Exl},
  \citenamefont {Bance}, \citenamefont {Reichel}, \citenamefont {Schrefl},
  \citenamefont {Peter~Stimming},\ and\ \citenamefont {Mauser}}]{Exl2014a}%
  \BibitemOpen
  \bibfield  {author} {\bibinfo {author} {\bibfnamefont {L.}~\bibnamefont
  {Exl}}, \bibinfo {author} {\bibfnamefont {S.}~\bibnamefont {Bance}}, \bibinfo
  {author} {\bibfnamefont {F.}~\bibnamefont {Reichel}}, \bibinfo {author}
  {\bibfnamefont {T.}~\bibnamefont {Schrefl}}, \bibinfo {author} {\bibfnamefont
  {H.}~\bibnamefont {Peter~Stimming}},\ and\ \bibinfo {author} {\bibfnamefont
  {N.~J.}\ \bibnamefont {Mauser}},\ }\href {https://doi.org/10.1063/1.4862839}
  {\bibfield  {journal} {\bibinfo  {journal} {Journal of Applied Physics}\
  }\textbf {\bibinfo {volume} {115}},\ \bibinfo {pages} {128} (\bibinfo {year}
  {2014})}\BibitemShut {NoStop}%
\bibitem [{\citenamefont {Cortes}(2017)}]{Cortes2017}%
  \BibitemOpen
  \bibfield  {author} {\bibinfo {author} {\bibfnamefont {D.~I.}\ \bibnamefont
  {Cortes}},\ }\emph {\bibinfo {title} {Computational Simulations of Complex
  Chiral Magnetic Structures}},\ \href@noop {} {Ph.D. thesis},\ \bibinfo
  {school} {University of Southampton} (\bibinfo {year} {2017})\BibitemShut
  {NoStop}%
\bibitem [{\citenamefont {Wang}\ \emph {et~al.}(2018)\citenamefont {Wang},
  \citenamefont {Zhang}, \citenamefont {A.Pepper}, \citenamefont {Mu},
  \citenamefont {Zhou},\ and\ \citenamefont {Fangohr}}]{Wang2018}%
  \BibitemOpen
  \bibfield  {author} {\bibinfo {author} {\bibfnamefont {W.}~\bibnamefont
  {Wang}}, \bibinfo {author} {\bibfnamefont {Z.}~\bibnamefont {Zhang}},
  \bibinfo {author} {\bibfnamefont {R.}~\bibnamefont {A.Pepper}}, \bibinfo
  {author} {\bibfnamefont {C.}~\bibnamefont {Mu}}, \bibinfo {author}
  {\bibfnamefont {Y.}~\bibnamefont {Zhou}},\ and\ \bibinfo {author}
  {\bibfnamefont {H.}~\bibnamefont {Fangohr}},\ }\href@noop {} {\bibfield
  {journal} {\bibinfo  {journal} {J. Phys. Condens. Matter}\ }\textbf {\bibinfo
  {volume} {30}},\ \bibinfo {pages} {015801} (\bibinfo {year}
  {2018})}\BibitemShut {NoStop}%
\bibitem [{\citenamefont {Garanin}(1996)}]{Garanin1996}%
  \BibitemOpen
  \bibfield  {author} {\bibinfo {author} {\bibfnamefont {D.~A.}\ \bibnamefont
  {Garanin}},\ }\href {https://doi.org/10.1103/PhysRevB.53.11593} {\bibfield
  {journal} {\bibinfo  {journal} {Physical Review B}\ }\textbf {\bibinfo
  {volume} {53}},\ \bibinfo {pages} {11593} (\bibinfo {year}
  {1996})}\BibitemShut {NoStop}%
\end{thebibliography}%

\end{document}